\documentclass[prb,aps,twocolumn,floats,nofootinbib,showpacs,superscriptaddress]{revtex4}

\usepackage{amsmath,amssymb,dsfont,graphicx,psfrag}
\usepackage{epsfig}
\usepackage{graphicx}

\def\beq{\begin{equation}}
\def\eeq{\end{equation}}
\def\beqr{\begin{eqnarray}}
\def\eeqr{\end{eqnarray}}

\def\bB{{\mathbf B}}
\def\bK{{\mathbf K}}

\def\bq{{\bf q}}

\def\bx{{\bf x}}
\def\a{{\alpha}}
\def\b{{\beta}}
\def\e{{\epsilon}}
\def\ua{{\uparrow}}
\def\da{{\downarrow}}
\def\half{\frac{1}{2}}
\def\trho{{\tilde\rho}}
\def\tea{{\tilde \epsilon_a}}
\def\tg{{\tilde g}}
\def\cE{{\cal E}}
\def\tcE{{\tilde{\cal E}}}
\def\tD{{\tilde D}}

\begin{document}
\title{Spin-Valley Coherent Phases of the $\nu=0$ Quantum Hall State in Bilayer Graphene}

\author{Ganpathy Murthy}
\affiliation{Department of Physics and Astronomy, University of Kentucky, Lexington KY 40506-0055, USA}

\author{Efrat Shimshoni}
\affiliation{Department of Physics, Bar-Ilan University, Ramat-Gan 52900, Israel}
\author{H.~A.~Fertig}
\affiliation{Department of Physics, Indiana University, Bloomington, IN 47405, USA}

\date{\today}
\begin{abstract}
Bilayer graphene (BLG) offers a rich platform for broken symmetry
states stabilized by interactions.  In this work we study the phase
diagram of BLG in the quantum Hall regime at filling factor $\nu=0$
within the Hartree-Fock approximation.  In the simplest
non-interacting situation this system has eight (nearly) degenerate
Landau levels near the Fermi energy, characterized by spin, valley,
and orbital quantum numbers.  We incorporate in our study two effects
not previously considered: ({\it i}) the nonperturbative effect of
trigonal warping in the single-particle Hamiltonian, and ({\it ii})
short-range SU(4) symmetry-breaking interactions that distinguish the
energetics of the orbitals.  We find within this model a rich set of
phases, including ferromagnetic, layer-polarized, canted
antiferromagnetic, Kekul\'e, a ``spin-valley entangled'' state, and a
``broken U(1) $\times$ U(1)'' phase.  This last state involves
independent spontaneous symmetry breaking in the layer and valley
degrees of freedom, and has not been previously identified.  We
present phase diagrams as a function of interlayer bias $D$ and
perpendicular magnetic field $B_{\perp}$ for various interaction and
Zeeman couplings, and discuss which are likely to be relevant to BLG
in recent measurements.  Experimental properties of the various phases
and transitions among them are also discussed.
\end{abstract}
\maketitle

\section{Introduction}
\label{section1}
Two-dimensional systems with discrete degrees of freedom in the
quantum Hall regime support a variety of possible broken symmetry
states, a phenomenon known as quantum Hall ferromagnetism
(QHF)\cite{QHFM}.  In this context graphene has presented itself as a
particularly exciting system, both in its monolayer and bilayer forms.
These systems differ from more conventional two dimensional electron
gases in supporting a $\nu=0$ quantized Hall effect, a consequence of
negative energy levels that are necessarily present in their
non-interacting spectra \cite{Gusynin_2005,McCann_2006}.  Moreover,
the presence of nearly-degenerate Landau levels (arising from internal
degrees of freedom such as spin, valley, and layer) near the Fermi
energy in undoped systems suggest that these systems offer a unique
platform for QHF physics \cite{Yang2006}.

In this work we study QHF in bilayer graphene (BLG) subject to
magnetic and electric fields.  In zero magnetic field, working in the
tight-binding model with nearest-neighbor hoppings only, the system
distinguishes itself from single layer graphene at the noninteracting
level in supporting two quadratic band touching (QBT) points, at the
$K$ and $K'$ points in the Brillouin zone, in contrast with monolayer
graphene which supports Dirac points at these locations.  When
undoped, the Fermi energy passes through these QBT's, opening the
possibility of many-body instabilities when interactions are included
in zero magnetic field
\cite{Nandkishore-Levitov_2010,Vafek-Yang_2010,Zhang-MacDonald_2010,Vafek_2010}
.  In the presence of a field, this system supports eight Landau
levels near the Fermi energy, offering a particularly rich set of
possibilities for groundstates with broken symmetries. These levels
arise from spin and valley quantum numbers, as well as orbital states
$n=0,1$ which are degenerate at any magnetic field in the simplest
models, when no electric field $D_{\perp}$ is applied perpendicular to
the system.

Previous studies of this system have focused on models which differ in
their choice of physical effects retained in the single-particle
Hamiltonian, and in how interactions are modeled.  Projection of the
long-range Coulomb interaction into this 8-fold manifold yields an
effective Hamiltonian with a layer-polarized state at large
$D_{\perp}$ and a ferromagnetic state at small $D_{\perp}$, with a
first order transition separating them \cite{Barlas_2008,Gorbar_2011}.
Distinguishing intra- and inter-layer Coulomb interactions, as well as
inclusion of particle-hole symmetry-breaking terms, leads to the
appearance of a state spontaneously breaking a U(1) symmetry
\cite{Lambert_2013,lukose_2016,knothe_2016,Jia_2017}.

Interactions in general are, however, more complicated than the
long-range Coulomb form, because at the microscopic scale they may
have lower symmetry (e.g., on-site Hubbard interactions).  Moreover,
short-range interactions have greater effect than expected based on
projection directly into the small set of Landau levels near the Fermi
energy, because they impact the energetics of the Landau levels below
them
\cite{Herbut_2007,Shizuya_2012,Roy_2014_1,Roy_2014,Feshami_2016,lukose_2016,knothe_2016,Jia_2017}.
An effective method for dealing with this, introduced by Kharitonov
\cite{kharitonov_bulk_monolayer,kharitonov_bulk_bilayer}, uses
phenomenological short-range interactions consistent with the
symmetries of the lattice, in principle incorporating renormalizations
from the Landau levels deep within the Dirac sea. In this study, we adopt this general
approach of effective interactions confined to the set of Landau
levels near zero energy.

Experimentally, evidence for phase transitions among states of
different broken symmetries has been accumulating.  Two-terminal
conductance experiments reveal quantized Hall states at low and high
$D_{\perp}$ at filling factor $\nu=0$, interrupted at intermediate
$D_{\perp}$ scales by a region where the transport gap vanishes
\cite{Weitz_2010,velasco_2012,maher_2013}, indicating a phase
transition between different quantized Hall states.  The value of
$D_{\perp}$ at which this transition occurs increases monotonically
with increasing $B_{\perp}$, the magnetic field component
perpendicular to the bilayer.  The high $D_{\perp}$ phase is rather
naturally identified with a layer polarized state, while the low
$D_{\perp}$ phase is largely thought to represent a canted
antiferromagnet (CAF) phase as was suggested in
Ref. \onlinecite{kharitonov_bulk_bilayer}.  More recent capacitance
measurements \cite{Hunt_2016}, however, show signatures of a separate
intermediate gapped phase between the low and high $D_{\perp}$ limits,
appearing above $B_{\perp} \sim$ 12T -13T.  Finally, in some samples
the region in $D_{\perp}$ separating the low and high $D_{\perp}$
states even at lower $B_{\perp}$ is not perfectly sharp, raising the
possibility of other phases existing in the transition
region\cite{maher_2013,jun-pvt}.

\begin{figure}
\includegraphics[width=1.0\linewidth]{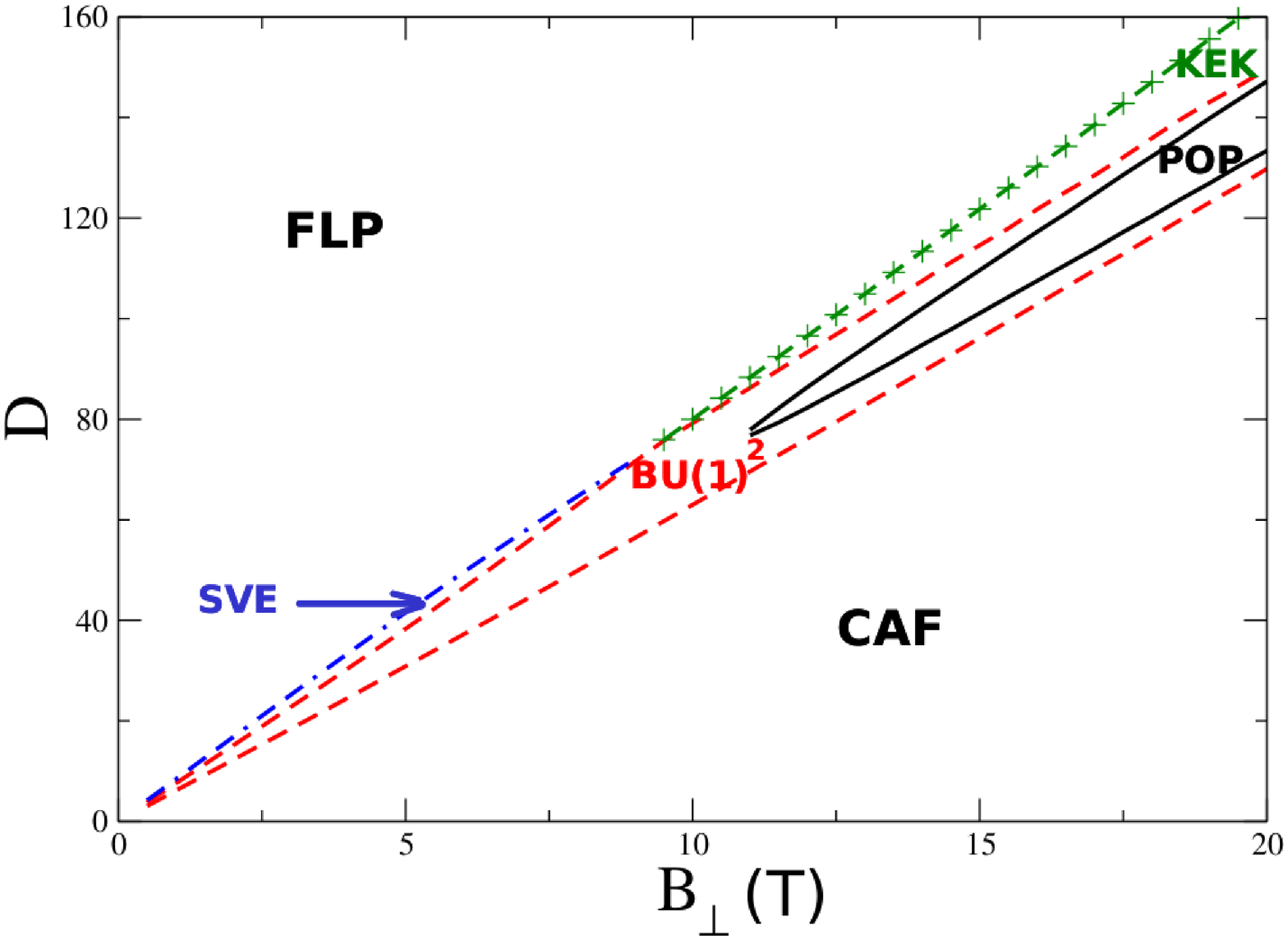}
\caption{The theoretical phase diagram in the tuning parameters
  $B_\perp$ and the perpendicular electric field (labelled $D$ in the
  figure and proportional to $D_\perp$) of our model in a range of
  assumed couplings which exhibits the Broken-U(1)$\times$U(1)
  (BU(1)$^2$) state. Here and in all the figures following, $B_\perp$
  is in Tesla, and $D$ is in arbitrary units. The boundaries of the
  BU(1)$^2$ state are the dashed red lines, while the boundaries of
  the partially orbitally polarized (POP) state are the solid black
  lines. The blue dash-dotted line is the upper boundary of the
  spin-valley entangled (SVE) phase, while the green dashed line with
  the + symbols is the upper boundary of the Kekule (KEK) state. The
  canted antiferromagnet (CAF) occupies the small $D$ part of the
  diagram at all values of $B_\perp$.  For small values of
  $B_\perp\leqslant11$T, as one increases $D$ starting from zero, one
  successively encounters the CAF state, the BU(1)$^2$ state, the
  spin-valley-entangled (SVE) state, and finally, at large values of
  $D$, the fully layer polarized (FLP) state. At larger values of
  $B_\perp>11$T, again starting from $D=0$, one encounters the CAF,
  the BU(1)$^2$ state, the partially orbitally polarized (POP) state,
  the Kekule (KEK) state, and finally the FLP state. All solid lines
  indicate first-order phase transitions while the broken lines
  indicate second-order transitions.}
\label{PS1-Fig1-bperp-phase-dia-fig}
\end{figure}

In this work, we explore the phase diagram of bilayer graphene at
$\nu=0$ using a model of the form introduced in
Ref. \onlinecite{kharitonov_bulk_bilayer}, within the Hartree-Fock
approximation.  Our model incorporates two ingredients which, to our
knowledge, have not been considered before in the context of
interacting BLG.  The first is the nonperturbative inclusion of
``trigonal warping'' \cite{McCann_2006} (arising from a hopping
amplitude $t_3$ between sites in different layers which are not above
one another) in the single particle states comprising the low-energy
manifold. Here and in the following, by ``low-energy manifold'' we
will mean the states lying near the Fermi energy. The $t_3$ term is
allowed by the spatial symmetries of the lattice, and generically
arises in {\it ab initio} approaches to the band structure of BLG (see
Ref. \onlinecite{Jung14} and references therein).  This hopping term
significantly distorts the QBT in zero field, replacing it with four
Dirac points \cite{McCann_2006}.  From a renormalization group (RG)
perspective, recent work \cite{Pujari_2016} has shown that the $t_3$
term, being allowed by symmetry, is generated by short-range
interactions, even if it is assumed to be zero in the bare
theory. Once generated, it is relevant, and flows to large values at
low energies.  In large magnetic fields this term has a very small
effect \cite{McCann_2006}. In consequence, this term has previously
been either neglected
\cite{kharitonov_bulk_bilayer,Lambert_2013,lukose_2016,Jia_2017} or
taken into account only perturbatively \cite{knothe_2016}.  We find,
however, that for experimentally relevant values of $B_{\perp}$ the
nonperturbative effect of the $t_3$ term is crucial to stabilizing
hitherto unknown broken symmetry states.

The second crucial element in our theory is the inclusion of short range
interactions not included in
Ref. \onlinecite{kharitonov_bulk_bilayer}: a density-density coupling
$g_0$, and an orbital anisotropy coupling $g_{nz}$, an Ising-like
interaction energy for fluctuations in the density differences between
the two spatial orbitals.  Both these couplings are allowed by
symmetry, and we find that including them yields a minimal model with a
phase diagram qualitatively consistent with current experimental
observations.

The phases that we find to be stable in different parameter regimes
include: (1) a fully layer polarized (FLP) state, (2) a fully spin
polarized (ferromagnetic, FM) state; (3) a canted antiferromagnetic
state (CAF), characterized by partial spin alignment along the
direction of the total magnetic field and antiferromagnetic alignment
between electrons in different valleys; (4) a Kekul\'e state (KEK),
which may be regarded as an analog of the CAF in which the roles of
spin and valley degrees of freedom have been interchanged; (5) a
``spin-valley entangled'' (SVE) state, in which the occupied
single-particle states involve coherent superpositions of states of
opposing spin and valley index, similar to the spin-layer coherent
state of Refs. \onlinecite{Lambert_2013,knothe_2016}; (6) a partial orbitally
polarized (POP) state; and finally (7) a more exotic
``Broken U(1)$\times$U(1)'' state, which supports non-trivial
coherence among different combinations of the single-particle states
in the spin-valley manifold such that two different U(1) symmetries
are spontaneously broken.  This contrasts with the other coherent
states that we find (which have been discussed in earlier literature
as well
\cite{kharitonov_bulk_bilayer,Lambert_2013,lukose_2016,knothe_2016,Jia_2017}) --
the CAF, KEK, and SVE -- which represent families of states with a
single spontaneously broken U(1) symmetry.

To our knowledge the Broken-U(1)$\times$U(1) (BU(1)$^2$) state has not
been previously identified in the literature, though hints of it have
been seen in the vanishing energy of collective modes even at $t_3=0$
\cite{denova_2017} at the CAF/FM to KEK/FLP phase boundary (we explain
this connection in Sections \ref{sve}, \ref{instabilities} and
\ref{largebperp}). Within our model, the BU(1)$^2$ phase requires a
nonzero trigonal warping in the single particle Hamiltonian, as well
as the $g_0$ and $g_{nz}$ couplings.  We find that for physically
reasonable sets of parameters it connects states with fewer broken
symmetries, such as the CAF and KEK as the interlayer potential
$D_{\perp}$ or the perpendicular field $B_{\perp}$ increases. Each of
the two U(1) angles involved comes with a stiffness, one or the other
of which vanishes continuously as the transition to another state is
approached.  This suggests the possibility of thermal or quantum
disordering of the phase, and the possibility that the state does not
manifest the quantized Hall effect at experimentally relevant
temperatures.  If so, this would introduce a broad transition region
between, for example, CAF and FLP states as a function of $D_{\perp}$,
rather than a sharp transition between them as would occur in a
first-order transition.  A typical phase diagram is illustrated in
Fig. \ref{PS1-Fig1-bperp-phase-dia-fig}.

The rest of this article is organized as follows. In Section
\ref{section2} we introduce the noninteracting Hamiltonian for BLG and
the low-energy basis states we will be using. These basis states
include the effect of the trigonal warping nonperturbatively.  In
Section \ref{IntHam} we will introduce the interacting Hamiltonian,
and present the general formula for the energy of a Hartree-Fock (HF)
state. In Section \ref{HF-and-Instabilities} we describe the states
that are encountered in our numerical calculation. We also present the
linear instabilities of these states which helps us identify various
second-order phase transitions. Most importantly, it helps us identify
three different regimes of the coupling constants which result in
different topologies of the phase diagram. In Section \ref{Results},
we present a brief analysis of the possible phase diagrams at small
$B_\perp$ and large $B_\perp$. This distinction arises because the
term in the Hamiltonian induced by the trigonal warping scales as
$\sqrt{B_\perp}$, whereas other terms are proportional to
$B_\perp$. Section \ref{Results} also contains our main results. These
include phase diagrams in $B_\perp - D$ space ($D$ is proportional to
the perpendicular electric field applied on the sample) for three
different regimes of coupling constants that produce different
topologies for the phase diagrams.  Section \ref{discussion} includes
a discussion of experimental consequences relevant to our phase
diagrams, and notes a few limitations of our analysis. Section
\ref{Conclusions} concludes with a summary, open questions, and future
directions.

\section{Noninteracting Hamiltonian and Low Energy States}
\label{section2}

To set our notation from the start, we will use the index $n=0,1$ for
the orbital degree of freedom, the Greek indices $\a=0,1$ for the
valley (where $\a,\beta=0\equiv K$ and $\a,\beta=1\equiv K'$), and the indices
$s,s'=0,1$ for spin ($s=0\equiv\uparrow$, and $s=1\equiv\downarrow$). As
a starting point for analyzing the single-body part of the Hamiltonian
we consider a Bernal stacked BLG, where the A site of one layer is
directly on top of the B$^\prime$ site of the other. In the presence
of a perpendicular electric field $D_\perp$ and a magnetic field ${\bf
  B}$ (introduced via a gauge choice where $A_y=B_\perp x$), the
approximate effective Hamiltonian describing electron states on the
remaining two sites of the BLG unit cell is given (for valley {\bf K},
spin $s=0,1=\uparrow,\downarrow$ and wave-vector $k$ in the
$\hat{y}$-direction) by \cite{McCann_2006,Jung14} \beqr H^{K
  s}_{eff}&=&H_0+H_Z+H_D \label{eq:Heff}\\ H_{0}&=&-\hbar\omega_c\left(\begin{array}{cc}
  -\tea a^\dagger a & \left(a^\dagger\right)^{2}+\lambda
  a\\ \left(a\right)^{2}+\lambda a^\dagger & -\tea aa^\dagger
\end{array}\right)\; ,\nonumber\\
H_Z&=&-\left(\begin{array}{cc} E_z(-1)^s & 0 \\ 0 & E_z(-1)^s \end{array}\right)\; ,\nonumber\\ H_D&=&-\left(\begin{array}{cc} D & 0 \\ 0 & -D \end{array}\right) \; .\nonumber
\label{jungMacDH}\eeqr
Here $E_z\propto |{\bf B}|$ is the Zeeman energy, $D\propto D_\perp$
is (half) the interlayer bias, and
$a=\frac{\ell}{\sqrt{2}}[\partial_x+(x-X)/\ell^2]$ is the Landau level
lowering operator (with $\ell=\sqrt{\hbar c/e B_\perp}$ the magnetic
length and $X=k\ell^2$ the guiding center coordinate). The parameters of $H_0$ account for all the
tight-binding parameters listed in Ref. \onlinecite{Jung14}, including
the longer-range interlayer hopping coefficients $t_3$, $t_4$ and a
particle-hole breaking onsite energy $\Delta$:
\beqr \omega_c &= & \frac{\hbar}{\ell^2
  m}\sim B_\perp \; ,\label{omega_c} \\ m &\equiv &
\frac{t_1^2-\Delta^2}{2[t_1(v_\perp^2+v_4^2)+2v_\perp
    v_4\Delta]}\approx \frac{t_1}{2v_\perp^2} \nonumber \eeqr where
$v_\perp=\sqrt{3}|t_\perp|a_0/2\hbar$, $v_4=\sqrt{3}|t_4|a_0/2\hbar$
with $t_\perp$ the inlayer hopping obeying $|t_\perp|\gg
|t_4|,|t_1|\gg\Delta$. The dimensionless parameter \beq
\tea=\frac{[\Delta(v_\perp^2+v_4^2)+2t_1v_\perp
    v_4]}{[t_1(v_\perp^2+v_4^2)+2v_\perp v_4\Delta]}\approx
\frac{\Delta}{t_1}
\label{tea_def}
\eeq
determines the orbital anisotropy energy, and is independent of $B_\perp$, whereas
\beq
\lambda=\frac{\sqrt{2}v_3m\ell}{\hbar}=\frac{\sqrt{3/2}|t_3|a_0m\ell}{\hbar^2}\sim\frac{1}{\sqrt{B_\perp}}\; .
\label{lambda_def}
\eeq
Finally, $H^{K^\prime s}_{eff}$ (for the other valley {\bf K$^{\mathbf\prime}$}) can be obtained from Eq. (\ref{eq:Heff}) by
trading $a^\dagger\leftrightarrow a$, $D\leftrightarrow -D$ and $\lambda\leftrightarrow -\lambda$.

The spectrum and eigenstates of the above effective Hamiltonian are
well-known for the case $\lambda=\tea=0$, i.e. when subleading hopping
parameters are neglected. In particular, there is a two-fold orbitally
degenerate manifold of zero energy eigenstates of $H_0$ (ignoring spin
and the guiding center indices for the moment):
\beq
|n,K\rangle=\left(\begin{array}{cc} |n\rangle \\ 0 \end{array}\right)\; ,\quad |n,K^\prime\rangle=\left(\begin{array}{cc} 0 \\ |n\rangle \end{array}\right),
\label{lambda0states}
\eeq
where $|n\rangle$ with $n=0,1$ are Landau level (LL)
wavefunctions. Their corresponding energies are
$\epsilon_{n,\a,s}=-D(-1)^\a-E_z(-1)^s$. Note that the two-fold
degeneracy of $n=0,1$ can be traced back to the quadratic
band-touching (QBT) characteristic to BLG. Adding a finite $\tea$ to
$H_0$ [Eq. (\ref{eq:Heff})] maintains the eigenstates
[Eq. (\ref{lambda0states})], and merely lifts the degeneracy of the
$n=0,1$ orbitals by a small asymmetry energy. However, the parameter
$\lambda$ associated with the $t_3$-hopping term, which introduces
trigonal warping of the QBT, fundamentally changes the structure of
the electronic states. Moreover, using empirical estimates of the bare
parameters \cite{Jung14,t4t3exp} in Eq. (\ref{lambda_def}), one
obtains $\lambda\equiv\lambda_1/\sqrt{B_\perp}$ where $B_\perp$ is in
Tesla and $\lambda_1\sim 1$ is the value of $\lambda$ at $B_\perp=1$
T. This implies that its effect is not necessarily perturbative; its
relative significance is tunable with $B_\perp$, and becomes
especially pronounced for moderately low fields of the order of a
Tesla. Indeed, as we show below, the resulting change in the structure
of non-interacting electron states has dramatic consequences on the
nature of broken-symmetry states when interactions are included.

We therefore focus on the case where $\lambda\neq 0$ is arbitrary, and
$\tea=0$ (corrections due to a finite $\tea$ will be accounted for
later on as a perturbation). The eigenstates of $H^{K s}_{eff}$,
$H^{K^\prime s}_{eff}$ can then be cast as (again ignoring spin and
guiding center indices)
\beqr
|K\rangle=\left(\begin{array}{cc} |\psi_{_K}\rangle \\ 0 \end{array}\right)\; ,&\quad &|K^\prime\rangle=\left(\begin{array}{cc} 0 \\ |\psi_{_{K'}}\rangle \end{array}\right) \nonumber\\ {\rm where}\quad (a^2+(-1)^\a\lambda a^\dagger)|\psi_\a\rangle &=& 0\; .
\label{psi_general}
\eeqr
Using the operator identity $[a,f(a^\dagger)]=f^\prime(a^\dagger)$ (with $f(x)$ an analytic function), Eq. (\ref{psi_general}) can be cast as an operator version of the Airy equation $y^{\prime\prime}-xy=0$ whose solutions are the functions\cite{ASbook} $Ai(x)$, $Bi(x)$. Employing their integral form, we obtain the following basis for the states $|\psi_{_K}\rangle$ (i.e., for $\alpha=0$ and $\lambda>0$):
\beqr
\label{psi_AB_integral}
|\psi_A,K\rangle &=&\int_0^\infty dt\left[\cos\left(\frac{t^3}{3\lambda}-ta^\dagger\right)\right]|0\rangle, \\
|\psi_B,K\rangle &=&\int_0^\infty dt\left[e^{-\frac{t^3}{3\lambda}-ta^\dagger}+\sin\left(\frac{t^3}{3\lambda}-ta^\dagger\right)\right]|0\rangle\; .\nonumber
\eeqr
It is convenient to express these integral forms as power series in $\lambda$. This yields $|\psi_A\rangle$, $|\psi_B\rangle$ as linear combinations of the orthonormal orbital states
(see Appendix \ref{Anm})
\begin{widetext}
\beqr
\label{psi_01_def}
|\psi_0,K\rangle &=&\sum_{m=0}^\infty A_{0m}|3m\rangle\; ,\quad A_{0m}\equiv C_0(-1)^m\frac{(3\lambda)^m}{\sqrt{(3m)!}}\frac{\Gamma(m+\frac{1}{3})}{\Gamma(\frac{1}{3})} ;\\
|\psi_1,K\rangle &=&\sum_{m=0}^\infty A_{1m}|3m+1\rangle\; ,\quad A_{1m}\equiv C_1(-1)^m\frac{(3\lambda)^m}{\sqrt{(3m+1)!}}\frac{\Gamma(m+\frac{2}{3})}{\Gamma(\frac{2}{3})}\; ,\nonumber
\eeqr
\end{widetext}
where $|N\rangle=\frac{1}{\sqrt{N!}}(a^\dagger)^N|0\rangle$ are the LL states and the normalization factors $C_n$ guarantee $\sum_0^\infty A_{nm}^2=1$.   Recalling Eq. (\ref{psi_general}), the solutions for the wavefunction $|\psi_{_{K'}}\rangle$ are directly obtained from Eq. (\ref{psi_01_def}) by the substitution $\lambda\rightarrow -\lambda$. For convenience, we recall our label $\alpha$ for the valleys such that $\alpha=K=0$, $\alpha=K'=1$, and the corresponding orbital labels $n=0,1$, so that
\beq
|n,\alpha\rangle\equiv\sum_{m=0}^\infty (-1)^{m\alpha}A_{nm}|3m+n\rangle\; .
\label{nalpha_def}
\eeq
The eigenstates of the effective Hamiltonian [with $\tilde{\epsilon}_{\alpha}=0$ in Eq. (\ref{eq:Heff})] are then given by
\beqr
|0,K,s\rangle=\left(\begin{array}{cc} |0,0,s\rangle \\ 0 \end{array}\right)\; ,&\quad & |1,K,s\rangle=\left(\begin{array}{cc} |1,0,s\rangle \\ 0 \end{array}\right)\; ,\nonumber\\
|0,K^\prime,s\rangle=\left(\begin{array}{cc} 0 \\ |0,1,s\rangle \end{array}\right)\; ,&\quad & |1,K^\prime,s\rangle=\left(\begin{array}{cc} 0 \\ |1,1,s\rangle \end{array}\right)
\label{eigenstates_final}
\eeqr
where the explicit dependence on the parameter $\lambda$ is given in Eqs. (\ref{psi_01_def}) and (\ref{nalpha_def})
and the states $|n,\alpha,s\rangle \equiv |n,\alpha\rangle \otimes |s\rangle$ incorporate spin.  Note that the wavevector $k$, or equivalently the guiding center $X=k\ell^2$,
is also a quantum number of the states, but is suppressed in the above expressions. 

This basis of low-energy states,
i.e., states close to the Fermi energy, has the full nonperturbative
dependence on $t_3$ which will turn out to be important for the rest
of our analysis.

To evaluate the energy spectrum, we consider the full effective Hamiltonian where the anisotropy parameter $\tea$ in Eq. (\ref{eq:Heff}) is finite but small [see Eq. (\ref{tea_def})], so that the corresponding terms can be treated perturbatively. Using the matrix elements
\beqr
\langle 0,\alpha|a^\dagger a|0,\alpha\rangle &=&\sum_{m=0}^\infty 3m|A_{0m}|^2, \\
\langle 1,\alpha|a^\dagger a|1,\alpha\rangle &=&\sum_{m=0}^\infty (3m+1)|A_{1m}|^2\; ,\nonumber
\eeqr
and implementing the substitution $D\rightarrow -D$ for $K\rightarrow K^\prime$, we obtain the energy levels corresponding to the states Eq. (\ref{eigenstates_final}) to first order in $\tilde{\epsilon}_{\alpha}$:
\beqr
\label{epsilon_nalpha}
\epsilon_{0,{_K}} &=& D+\tilde{\epsilon}_{\alpha}\sum_{m=0}^\infty 3m|A_{0m}|^2, \\ \epsilon_{1,{_K}} &=& D+\tilde{\epsilon}_{\alpha}\sum_{m=0}^\infty (3m+1)|A_{1m}|^2, \nonumber\\
\epsilon_{0,{_{K^\prime}}} &=& -D+\tilde{\epsilon}_{\alpha}\sum_{m=0}^\infty 3m|A_{0m}|^2, \nonumber\\ \epsilon_{1,{_{K^\prime}}} &=& -D+\tilde{\epsilon}_{\alpha}\sum_{m=0}^\infty (3m+1)|A_{1m}|^2 \; .\nonumber
\eeqr
For each valley, this introduces an orbital anisotropy
\beq
\epsilon_a\equiv\epsilon_{1,\alpha}-\epsilon_{0,\alpha}=\tilde{\epsilon}_{\alpha}\sum_{m=0}^\infty \left[(3m+1)|A_{1m}|^2-3m|A_{0m}|^2\right]
\label{epsilon_a_def}
\eeq
which can be numerically evaluated for arbitrarily large $\lambda$ using the expressions for $A_{nm}$ [Eq. (\ref{psi_01_def})].

\section{The Interaction Hamiltonian and Hartree-Fock}
\label{IntHam}

As explained above, there are three discrete quantum numbers for the
non-interacting single particle states in BLG, representing spin,
valley, and the $n=0,\ 1$ orbitals. To begin dealing with interactions
we divide the basic Coulomb interaction into a long-range part that
has the full SU(4) symmetry of spin and valley indices, and an
effective short-range part. The short-range interactions (including
those present at the bare level) should have SU(2) symmetry in the
spin sector and a U(1) symmetry in the valley sector. There is no
symmetry constraint in the orbital sector. Upon the application of a
Zeeman field the symmetry of the spin-sector is also reduced to a
U(1). Thus the symmetry of the full Hamiltonian is
U(1)$_{spin}\times$U(1)$_{valley}$.

Following previous work in
single layer graphene\cite{kharitonov_bulk_monolayer}, we will assume
that the relevant interactions at low energy have no explicit
spin-dependence. Translation invariance implies that at low energy
there should be two kinds of interactions, those that transfer a
momentum small compared to a reciprocal lattice vector, and those that
transfer a momentum close to the intervalley momentum
$\Delta{\bK}={\bK}-{\bK}'$. Taking all these conditions into account,
we obtain a large set of possible interactions, each with its own
coupling.

Such a high-dimensional coupling constant space is very hard to
analyze systematically. Hence, in this work, we will simplify the
system by considering a ``minimal'' model which contains only four
distinct couplings.  Defining $c_{n\a s k}$ as the destruction
operator for a particle in a $|n,\alpha,s,k\rangle$ state (here $k$ is
the Landau guiding center label), our minimal interaction Hamiltonian takes
the form
\begin{widetext}
\beqr
H_{int}=&\frac{1}{2L_xL_y}\sum\limits_{k_1,k_2,\bq} e^{-iq_x(k_1-k_2-q_y)\ell^2}\times\nonumber\\
&\bigg(v_0(\bq)\sum\limits_{n_im_i\alpha\beta
  s_1s_2}\trho_{n_1n_2}^{\alpha\alpha}(\bq)\trho_{m_1m_2}^{\beta\beta}(-\bq):{c}^{\dagger}_{n_1\alpha s_1,k_1-q_y }{c}_{n_2\alpha s_1,k_1}{c}^{\dagger}_{m_1\beta s_2,k_2+q_y }{c}_{m_2\beta s_2,k_2 }:\nonumber\\
&+ v_z(\bq)\sum\limits_{n_im_i\alpha\beta
  s_1s_2}\trho_{n_1n_2}^{\alpha\alpha}(\bq)\trho_{m_1m_2}^{\beta\beta}(-\bq):{c}^{\dagger}_{n_1\alpha s_1,k_1-q_y }\tau_z{c}_{n_2\alpha s_1,k_1 }{c}^{\dagger}_{m_1\beta s_2, k_2+q_y}\tau_z{c}_{m_2\beta s_2, k_2}:\nonumber\\
&+ 2v_{xy}(\bq)\sum\limits_{n_im_i s_1 s_2}\trho_{n_1n_2}^{KK'}(\bq)\trho_{m_1m_2}^{K'K}(-\bq):{c}^{\dagger}_{n_1K s_1, k_1-q_y}{c}_{n_2K' s_1, k_1}{c}^{\dagger}_{m_1K' s_2, k_2+q_y}{c}_{m_2K s_2, k_2}:\nonumber\\
&+v_{nz}(\bq)\sum\limits_{n_1n_2\alpha\beta s_1s_2}(-1)^{n_1+n_2}\trho_{n_1n_1}^{\alpha\alpha}(\bq)\trho_{n_2n_2}^{\beta\beta}(-\bq):{c}^{\dagger}_{n_1\alpha s_1, k_1-q_y}{c}_{n_1\alpha s_1, k_1}{c}^{\dagger}_{n_2\beta s_2, k_2+q_y}{c}_{n_2\beta s_2, k_2}:\bigg).
\eeqr
\end{widetext}
The matrix elements of the density $\tilde\rho_{n_1n_2}^{\alpha\beta}$
are defined using the states of Eq. (\ref{nalpha_def}) (with spin
still suppressed but the guiding center indices now explicit) as
\beq
\langle n_1\alpha k_1|e^{-i\bq\cdot\bx}|n_2\beta k_2\rangle=\delta_{k_1,k_2-q_y}e^{-iq_x(k_1-q_y/2)}\tilde\rho_{n_1n_2}^{\alpha\beta}(\bq).
\label{eq:trho}
\eeq
Some details about these matrix elements that are relevant to our
study are provided in Appendix \ref{app:FF}.  The couplings $v_z,\ v_{xy}$ were
originally introduced by Kharitonov for monolayer graphene\cite{kharitonov_bulk_monolayer}, and have
exactly the same meaning here as in the monolayer. In earlier work on
the edge states of monolayer graphene \cite{us_2014,us_2016}, we introduced the coupling
$v_0$, which treats all the discrete labels equally and endows the
system with a spin stiffness for spatial variations of the order
parameter. The new coupling we introduce is $v_{nz}$, which is
analogous to $v_z$, but in the orbital sector.

To proceed one must specify
forms for $v_0(\bq),\ v_z(\bq),\ v_{xy}(\bq)$, and $v_{nz}(\bq)$. We
make the simplest possible choices, that they are constants
independent of $\bq$. This means the interactions are very
short-ranged in space. We note that in the case of single-layer
graphene $v_0$ does not alter the relative energies of the various
possible bulk states. However, as we will see shortly, in bilayer
graphene $v_0$ enters the energies of different states with different
coefficients, and hence plays a role in picking the true ground state.

The full effective Hamiltonian of our system truncated to the
low-energy space is $H_0+H_{int}$ where

\beq
H_0=-\sum\limits_{n \alpha s k}{c}^{\dagger}_{n\alpha s k}{c}_{n\alpha s k}\big[(-1)^n\e_a+(-1)^sE_Z+(-1)^\a D\big].
\eeq
Any Hartree-Fock (HF) state is fully determined by its one-body
averages $\langle{c}^{\dagger}_i{c}_j\rangle$. We only consider states
in the bulk that conserve the guiding center label $k$: Thus, the only
possible translation symmetry breaking could arise via densities with
momenta ${\bK}-{\bK}'$. We define the matrix
$\Delta_{mn;ss'}^{\alpha\beta}$ via
\beq
\langle HF|{c}^{\dagger}_{m\alpha s k}{c}_{n\beta s' k'}|HF\rangle \equiv \delta_{kk'}\Delta_{mn;ss'}^{\alpha\beta}
\eeq
where $|HF\rangle$ is a Hartree-Fock state.
Note that $\Delta$ is independent of $k$.
Now consider evaluating the average of $H_{int}$ in such a state. A
generic term is a sum of direct and exchange contributions -- i.e.,
\begin{widetext}
\beq
\langle HF| {c}^{\dagger}_{n_1\alpha s_1, k_1-q_y}{c}^{\dagger}_{m_1\eta s_2, k_2+q_y}{c}_{m_2\gamma s_2, k_2}{c}_{n_2\beta s_1,k_1 }|HF\rangle=\delta_{q_y,0}\Delta_{n_1n_2;s_1s_1}^{\alpha\beta}\Delta_{m_1m_2;s_2s_2}^{\eta\gamma} -\delta_{k_1,k_2+q_y}\Delta_{n_1m_2;s_1s_2}^{\alpha\gamma}\Delta_{m_1n_2;s_2s_1}^{\eta\beta}.
\eeq
\end{widetext}
The direct terms are easy to deal with because
$\trho_{n_1n_2}^{\alpha\beta}(\bq=0)=\delta_{n_1n_2}\delta_{\alpha\beta}$. The
exchange integrals are a bit more involved. In Appendix \ref{app:FF} we show the
following important result, which is relevant because of our assumption that
all interactions $v_i(\bq)$ are constants $v_i$:
\beqr
\int\frac{d^2q}{(2\pi)^2}\trho_{n_1n_2}^{\alpha\beta}(\bq)\trho_{m_1m_2}^{\eta\gamma}(-\bq)=&\frac{\delta_{n_1m_2}\delta_{m_1n_2}}{2\pi\ell^2}  r_{\alpha\gamma}^{(n_1)}r_{\beta\eta}^{(n_2)},\nonumber\\
r_{\alpha\beta}^{(n)}=\sum\limits_{j=0}^{\infty} (-1)^{j(\alpha+\beta)} |A_{nj}|^2&= \left\{\begin{array}{cc}
  1&\a=\b\\
  r&\a\ne\b
\end{array}\right\} \\
{\rm where}\quad r = \sum_{k=0}^{\infty} (-1)^{k} A_{nk}^2. \nonumber
\label{eq:trho_integral}
\eeqr
The number $r$ is independent of the orbital index $n$ (see App. B) but does depend
on $B_{\perp}$ via the coefficient $\lambda$ [Eqs. (\ref{lambda_def}) and (\ref{psi_01_def})]
arising originally from the trigonal warping term $t_3$. The most
important consequence of this relation is that only $\Delta$'s diagonal
in the $n$-labels appear in the energy. Using the general reasoning of
Ref. \onlinecite{jungwirth_2000}, since the inter-orbital exchange (zero
here) is smaller than the intra-orbital exchange, this falls into the
Ising anisotropy class: The system cannot lower its energy by
superposing different orbitals in a single-particle
state. Operationally, this leads to the enormous simplification that
we need to consider only forms of $\Delta$ which are block-diagonal in
$n$:
\beq
\Delta_{n_1n_2;s_1s_2}^{\alpha\beta}\equiv \delta_{n_1n_2}\Delta_{n_1;s_1s_2}^{\alpha\beta}
\eeq

Let us now define the couplings $g_i=\frac{v_i}{2\pi\ell^2}$, and the
number of flux quanta passing through the sample
$N_\phi=\frac{L_xL_y}{2\pi\ell^2}$.
Recalling the indexing of Section \ref{section2}
($\alpha=0$ for the $K$ valley and 1 for the $K'$ valley,
$s=0$ for spin  up and 1 for spin down), the HF energy may then be written compactly as
\begin{widetext}
\beqr
\frac{\cE(\{\Delta\})}{N_\phi}\equiv\tcE(\{\Delta\})=&-\sum\limits_{n\alpha
  s}(\e_a(-1)^n+E_z(-1)^s+D(-1)^\alpha)\Delta_{n;ss}^{\alpha\alpha}\nonumber \\
&+\frac{g_0}{2}\bigg(\big(\sum\limits_{n\alpha s}\Delta_{n;ss}^{\alpha\alpha}\big)^2-\sum\limits_{\alpha\beta s s'}r_{\alpha\beta}^2\big(\sum_{n}\Delta_{n;ss'}^{\alpha\beta}\big)\big(\sum_{n'}\Delta_{n';s's}^{\beta\alpha}\big)\bigg)\nonumber\\
&+\frac{g_z}{2}\bigg(\big(\sum\limits_{n\alpha s}(-1)^n\Delta_{n;ss}^{\alpha\alpha}\big)^2-\sum\limits_{\alpha\beta s s'}r_{\alpha\beta}^2(-1)^{\alpha+\beta}\big(\sum_{n}\Delta_{n;ss'}^{\alpha\beta}\big)\big(\sum_{n'}\Delta_{n';s's}^{\beta\alpha}\big)\bigg)\nonumber\\
&+g_{xy}\bigg(r^2|\sum_{ns}\Delta_{n;ss}^{KK'}|^2-\sum_{ss'}\big(\sum_{n}\Delta_{n;ss'}^{KK}\big)\big(\sum_{n'}\Delta_{n',s's}^{K'K'}\big)\bigg)\nonumber\\
&+\frac{g_{nz}}{2}\bigg(\big(\sum\limits_{n\alpha s}(-1)^n\Delta_{n:ss}^{\alpha\alpha}\big)^2-\sum\limits_{nss'\alpha\beta}r_{\alpha\beta}^2\Delta_{n:ss'}^{\alpha\beta}\Delta_{n:s's}^{\beta\alpha}\bigg).
\eeqr
\end{widetext}

\section{Hartree-Fock States and Linear Instabilities}
\label{HF-and-Instabilities}

Before we present the numerical results, let us explore the nature of
the states we will encounter, parametrize them analytically, and find
critical values of $D$ at which one kind of state is unstable to
another. At $\nu=0$ four single-particle states must be filled at each
guiding center. All the states we consider are one of three types. (i) All
four occupied states could be in the same ($n=0$) orbital, which would
be a maximally orbitally anisotropic (MOA) state. (ii) Three of the occupied
states could be in the $n=0$ orbital while one is in the $n=1$
orbital, a partially orbitally polarized (POP) state. (iii)
Both the $n=0,\ 1$ orbitals support two occupied states. In this case
the most natural choice is
$\Delta_{0:ss'}^{\alpha\beta}=\Delta_{1:ss'}^{\alpha\beta}$, a state
symmetric in the orbital label. We will analyze each of these
possibilities in turn. In the following, when we represent
$\Delta_0$ and $\Delta_1$ as $4\times4$ matrices, our ordering will be
$K\ua,\ K\da,\ K'\ua,\ K'\da$. We will be guided by experiment in
choosing our parameters; in particular, we will consider only
$g_{xy}<0$, because of the evidence that a canted antiferromagnet
(CAF) state is stable in BLG,
determining the sign of $g_{xy}$.

\subsection{Maximally Orbitally Anisotropic State}
\label{moa}
This state is particularly simple. The $\Delta$ matrices are
\beq
\Delta_{0}=1_{4\times4},\ \ \ \ \ \Delta_1=0_{4\times4}.
\eeq
This state has orbital polarization, but no valley or spin
polarization. The HF energy is
\beq
\tcE_{MOA}=-4\e_a+6g_0-2g_z+6g_{nz}.
\eeq
We find, for our choices of parameters, that
this state is never the ground state.

\subsection{Partially Orbitally Polarized States}
\label{pop}

This state can be characterized by two different single-particle
states in the spin-valley sector, which for the moment
we generically label $|a\rangle$ and
$|b\rangle$. The $\Delta$ matrices can be described as
\beq
\Delta_0=1_{4\times4}-|a\rangle\langle a|,\ \ \ \ \Delta_1=|b\rangle\langle b|.
\eeq
In principle, the states $|a\rangle,\ |b\rangle$ can be arbitrary, but
at the HF minimum we find them to be parametrized by a single angle
$\theta$
\beqr
|a\rangle=&[\begin{array}{cccc}
    0&0&-\sin(\theta/2)&\cos(\theta/2)\\
    \end{array}]^T,\\
|b\rangle=&[\begin{array}{cccc}
    \cos(\theta/2)&-\sin(\theta/2)&0&0\\
\end{array}]^T,
\eeqr
where $\cos(\theta)=\frac{E_Z}{|g_{xy}|}$ for $E_Z<|g_{xy}|$ and
$\cos(\theta)=1$ for $E_Z>|g_{xy}|$. The energy of this state is
\beqr
\tcE_{POP}=&-2\e_a-2D+5g_0-g_z+|g_{xy}|-&\frac{E_Z^2}{|g_{xy}|}\\
&&E_Z<|g_{xy}|,\nonumber\\
=&-2\e_a-2D-2E_Z+5g_0-g_z+2|g_{xy}|&\\
&&E_Z>|g_{xy}|.\nonumber
\eeqr
Note that the POP states have an orbital polarization of 2, a valley
polarization of 2, and variable spin polarization which can never
exceed 2. They also spontaneously break the U(1) spin-rotation
symmetry around the direction of $\bB$ for $E_Z<|g_{xy}|$.

\subsection{States Symmetric in Orbitals}
\label{symmstates}

This class exhibits the richest set of HF states, and contains: (i)
The canted antiferromagnet (CAF) which spontaneously breaks the U(1)
spin-rotation symmetry around the direction of the total field
$\bB$. The fully spin-polarized ferromagnet (FM) is a limit of the
CAF. (ii) The Kekule state (KEK) which is a spin singlet but is canted
in the valley sector and thus spontaneously breaks the valley U(1)
symmetry. The fully layer polarized (FLP) state is a limit of the
Kekule state. (iii) A spin-valley-entangled (SVE) state that entangles
$K\da$ with $K'\ua$.  (iv) A new state which is canted in both the
spin and valley sectors, and thus has two distinct spontaneously
broken U(1) symmetries. We will call this state the
Broken-U(1)$\times$U(1), or BU(1)$^2$  state.

It will prove convenient to look at the $4\times4$ matrix
$\Delta_0=\Delta_1=\Delta$ rather than the occupied states themselves.
In all the orbitally symmetric states $g_{nz}$ only appears via the
combination $g_0+\half g_{nz}$. For future convenience we define
\beqr
&G_0=g_0+\half g_{nz},\\
\tD=&(1-r^2)G_0+(1+r^2)g_z+|g_{xy}|.
\label{G0Dtilde}\eeqr

\subsubsection{Canted Antiferromagnet (CAF) and Ferromagnet (FM)}
\label{caf/fm}

These states have a $\Delta$ matrix of the form
\beq
\Delta=\frac{1}{2}\left(\begin{array}{cccc}
  1+\cos\theta&\sin\theta&0&0\\
\sin\theta&1-\cos\theta&0&0\\
0&0&1+\cos\theta&-\sin\theta\\
0&0&-\sin\theta&1-\cos\theta\\
\end{array}\right).
\eeq
The minimum occurs at $\cos\theta=\frac{E_Z}{2|g_{xy}|}$ for $E_z \le 2|g_{xy}|$
and $\cos\theta=1$ for $E_z > 2|g_{xy}|$.  The energy is
\beqr
\tcE_{CAF}=&8g_0-4G_0-4g_z-\frac{E_Z^2}{|g_{xy}|}&\label{tcecaf}\\
&&E_Z\le 2|g_{xy}|,\nonumber\\
\tcE_{FM}=&8g_0-4G_0-4g_z-4E_Z+&4|g_{xy}|\label{tcefm}\\
&&E_Z>2|g_{xy}|.\nonumber
\eeqr
The case $E_Z>2|g_{xy}|$ corresponds to the fully spin-polarized FM
state. The CAF/FM state has only spin-polarization, and no orbital or
valley polarization. The CAF state spontaneously breaks the U(1)
spin-rotation symmetry around $\bB$. The FM state has no spontaneously broken symmetries.

\subsubsection{Kekule (KEK) and Fully Layer Polarized (FLP) States}
\label{kek/flp}

For this state,
\beq
\Delta=\frac{1}{2}\left(\begin{array}{cccc}
  1+\cos\theta&0&\sin\theta&0\\
  0&1+\cos\theta&0&\sin\theta\\
\sin\theta&0&1-\cos\theta&0\\
0&\sin\theta&0&1-\cos\theta\\
\end{array}\right).
\eeq
To specify the angle at the minimum, we need to define an energy $g_K$;
\beq
g_K=(3-r^2)g_z+(2r^2-1)|g_{xy}|-(1-r^2)G_0.
\label{gK}\eeq
In terms of $g_K$ the energy for arbitrary $\theta$ can be expressed as
\beqr
\tcE(\theta)=&8g_0-2(1+r^2)G_0-2(1-r^2)g_z\nonumber\\
&-2(2r^2-1)|g_{xy}|-4D\cos\theta+2g_K\cos^2\theta.
\eeqr
It is clear that if $g_K<0$, $\theta=0$ is the minimum. For $g_K>0$ we find $\theta$ at the minimum to be
\beq
\cos\theta=\frac{D}{g_K}\ \ \  \forall D<g_K;\ \ \ \ \cos\theta=1\ \ \  \forall D>g_K.
\eeq
The case $D>g_K$ corresponds to the fully layer polarized (FLP) state. The energy of the KEK/FLP state is
\beqr
\tcE_{KEK}=&8g_0-2(1+r^2)G_0-2(1-r^2)g_z\nonumber\\
&-2(2r^2-1)|g_{xy}|-\frac{2D^2}{g_K}\ \ \ \ D<g_K,\label{tcekek}\\
\tcE_{FLP}=&8g_0-4G_0+4g_z-4D\ \ \ \ D>g_K.\label{tceflp}
\eeqr
The KEK/FLP states have no orbital or spin polarization. They do have
a valley polarization. The KEK state spontaneously breaks the valley U(1)
symmetry. The FLP state does not spontaneously break any symmetry.

\subsubsection{Spin-Valley Entangled (SVE) State}
\label{sve}

This state has the $K\ua$ state occupied, but mixes the $K\da$ and
$K'\ua$ states. In this case,
\beq
\Delta=\frac{1}{2}\left(\begin{array}{cccc}
  2&0&0&0\\
  0&1+\cos\psi&\sin\psi&0\\
0&\sin\psi&1-\cos\psi&0\\
0&0&0&0\\
\end{array}\right).
\eeq
The energy of this state is evaluated to be
\beqr
\tcE(\psi)=&4(2g_0-G_0-g_z+|g_{xy}|-E_Z)-\nonumber\\
&4\cos^2\frac{\psi}{2}(D+2|g_{xy}|-E_Z-\tD)\nonumber\\
&+4(1-r^2)(g_z-G_0)\cos^4\frac{\psi}{2}.
\label{Esve}\eeqr
The optimum value of $\cos^2\frac{\psi}{2}$ is easily found to be
\beq
\cos^2\frac{\psi}{2}=\frac{D+2|g_{xy}|-\tD-E_Z}{2(1-r^2)(g_z-G_0)}.
\eeq
Defining
\beqr
&D^{SVE}_{min}=\tD+E_Z-2|g_{xy}|,\\
&D^{SVE}_{max}=D^{SVE}_{min}+2(1-r^2)(g_z-G_0),
\label{sveminmax}\eeqr
the minimum energy
of the SVE state for $D$ in the range $D^{SVE}_{min}< D < D^{SVE}_{max}$ is
\beq
\tcE_{SVE}=4\big(2g_0-G_0-g_z-E_Z+|g_{xy}|\big)-\frac{2(D-D^{SVE}_{min})^2}{(D^{SVE}_{max}-D^{SVE}_{min})}.
\eeq
This state spontaneously breaks a single U(1), which is an entangled
combination of valley and spin, and smoothly interpolates between the
FLP and the FM states.

Note that as $r^2\to1$, Eq. (\ref{sveminmax}) implies that the range
of $D$ over which the SVE state exists shrinks to zero. In fact,
precisely at $r^2=1$ and $D=D^{SVE}_{min}=D^{SVE}_{max}$ the energy of
Eq. (\ref{Esve}) becomes independent of $\psi$. This means that there
should be a zero energy $q=0$ collective mode at this value of $D$,
which is indeed seen in a recent calculation \cite{denova_2017}. This
is a hint of the potential existence of the SVE state {\it even at $r^2=1$}. 

\subsubsection{Broken U(1)$\times$U(1) [BU(1)\,$^2$] State}
\label{csvaf}

This is an interesting state that spontaneously breaks the U(1)
symmetries of both the spin and valley sectors. We will call this the
BU(1)$^2$ state for short. The most general state for two filled
levels, assuming real vectors, can be described by five real
parameters. This can be seen as follows: The first filled state is an
O(4) vector (real state) which can be specified by three angles. The
second filled state also has three angles, but the constraint that it should
be orthogonal to the first filled state reduces the total number of
independent angles by one, to a total of five.

We have numerically searched in this five-dimensional parameter space for the
minimum energy HF state, and found that these minima can always be described by
a state requiring only three real angles, which we call
$\theta,\ \chi,\ \psi$.
In addition to these
there are two $U(1)$ angles upon which the energy
does not depend, which we label $\phi$ and $\eta$.  Defining
$\gamma=\frac{\psi+\chi}{2}$ and $\zeta=\frac{\psi-\chi}{2}$ the resulting
$\Delta$ matrix may be expressed as
\begin{widetext}
\beq
\Delta=\frac{1}{2}\left(\begin{array}{cccc}
  1+\cos\chi\cos\theta&e^{-i\phi}\sin\theta\cos\zeta&e^{i\eta}\sin\theta\sin\zeta&-e^{i(\eta-\phi)}\sin\chi\cos\theta\\
e^{i\phi}\sin\theta\cos\zeta&1-\cos\psi\cos\theta&-e^{i(\eta+\phi)}\sin\psi\cos\theta&e^{i\eta}\sin\theta\sin\zeta\\
e^{-i\eta}\sin\theta\sin\zeta&-e^{-i(\eta+\phi)}\sin\psi\cos\theta&1+\cos\psi\cos\theta&-e^{-i\phi}\sin\theta\cos\zeta\\
-e^{-i(\eta-\phi)}\sin\chi\cos\theta&e^{-i\eta}\sin\theta\sin\zeta&-e^{i\phi}\sin\theta\cos\zeta&1-\cos\chi\cos\theta\\
\end{array}\right).
\label{FullDelta}
\eeq
\end{widetext}
The values of $\phi,\ \eta$ will be chosen in the true ground state by
spontaneous symmetry breaking. In the limit $\psi=\chi=0$ this ansatz
reduces to the CAF/FM where $\theta$ is the canting angle of the
CAF/FM. Similarly, for $\chi=0,\ \psi=\pi$ it reduces to the KEK/FLP
state, where $\theta$ now means the canting angle of the Kekule
state. Thus Eq. (\ref{FullDelta}) interpolates smoothly between the
CAF/FM and the KEK/FLP states. Finally, $\theta=\chi=\pi$ and $\psi
\ne 0,\pi$ corresponds to the SVE state. We will reserve the name
``Broken-U(1)$\times$U(1)'' for the state where all three angles
$\theta,\ \chi,\ \psi$ are nontrivial, that is, different from $0$ or
$\pi$.

The energy for this ansatz is
\begin{widetext}
\beqr
&\tcE=8g_0-4G_0-4g_z-4\cos\theta(E_Z\cos\gamma\cos\zeta+D\sin\gamma\sin\zeta)
+2\sin^2\zeta\big(\tD-2r^2|g_{xy}|\big)\nonumber\\
&+\cos^2\theta\big(4|g_{xy}|-4(1-r^2)|g_{xy}|\sin^2\zeta+2\sin^2\zeta[\tD-2|g_{xy}|]
+4\sin^2\gamma\sin^2\zeta(1-r^2)[g_z-G_0]\big).
\label{energy3angle}
\eeqr
\end{widetext}
Unfortunately, we have not been able to analytically find the minima
of $\tcE$ within its full three angle domain.

\subsection{Instabilities of CAF/FM, KEK/FLP, and SVE states}
\label{instabilities}

Since the three-angle ansatz can describe all the other states that
only have a single broken U(1), we can use the three-angle ansatz to
find the instabilities of the CAF/FM and the KEK/FLP. Motivated by
experiment, we will analyze the situation where $B_{\perp}$ and $E_Z$
are fixed while $D$ is varied. We define the critical $D$ at which the
CAF/FM becomes unstable to the three-angle ansatz as $D_{c1}$, while
the $D$ at which the KEK/FLP or the SVE/FLP becomes unstable to the
three-angle ansatz is defined as $D_{c2}$. Ignoring the POP state for
the moment, a necessary (but not sufficient) condition for the
Broken-U(1)$\times$U(1) state to exist as a HF state is
$D_{c2}>D_{c1}$.

\begin{figure}
\includegraphics[width=1.0\linewidth]{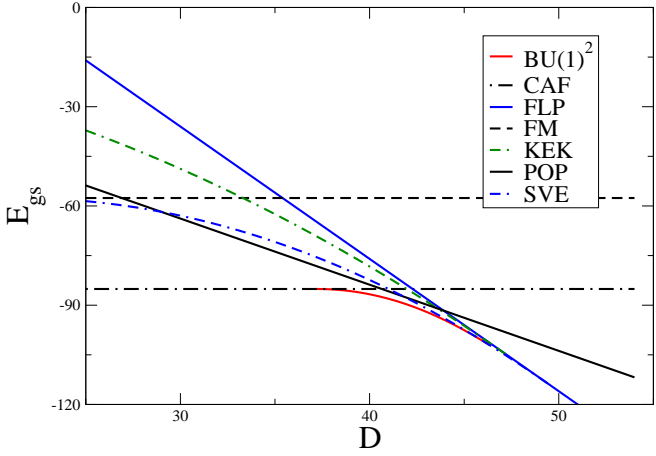}
\caption{The energies of the various HF states for a particular choice
  of parameters. We have fixed $B_\perp=6$ T, and $E_Z=|g_{xy}|/3$,
  and plotted the energies as functions of $D$. The dashed and the
  dot-dashed black lines are the FM and CAF energies
  respectively. They are independent of $D$. The solid blue line is
  the energy of the FLP state (with a slope of -4), while the solid
  black line is the energy of the POP state (with a slope of -2). The
  solid red line, where it exists, marks the energy of the BU(1)$^2$
  state. The dot-double-dashed blue line is the energy of the SVE
  state, which interpolates smoothly between the FM and FLP states.
  The dot-double-dashed green line is the energy of the KEK state. For
  this particular choice of parameters, as $D$ increases, the system
  starts in the CAF state for small $D$, undergoes a second-order
  transition into the BU(1)$^2$ state, which then gives way to the SVE
  state, which in turn yields to the FLP state. All transitions are
  second-order.}
\label{Energies-PS1-bperp6p0-fig}
\end{figure}

To make the ideas concrete, Fig. \ref{Energies-PS1-bperp6p0-fig} shows
the energies of the various HF states as functions of $D$ for fixed
$B_\perp=6$ T, and $E_Z=|g_{xy}|/3$. For this set of parameters, the
FM, KEK, and POP states are always higher in energy than the others,
and hence are not the ground state at any $D$. On the other hand, the
CAF, the BU(1)$^2$, the SVE, and the FLP states are the lowest in
energy, each in a corresponding range of $D$. As can be seen, the SVE
state interpolates smoothly between the FM and the FLP states. Thus,
the FLP and FM states must be linearly unstable to the SVE state at
the appropriate values of $D$. Similarly, the BU(1)$^2$ state
interpolates smoothly between the CAF and SVE states. Thus, the CAF
and SVE states must be linearly unstable to the BU(1)$^2$ state at the
appropriate values of $D$. In the following, we will analytically
compute the values of $D$ corresponding to the various linear
instabilities.

Let us consider $D_{c1}$ first. Since the CAF/FM state has
$\chi=\psi=\gamma=\zeta=0$, we can consider $\gamma,\ \zeta\ll 1$
and expand the energy in powers of $\gamma,\ \zeta$. After doing so,
we obtain a constant piece (the energy of the CAF/FM state) and a
quadratic form in $\gamma$ and $\zeta$. The instability occurs when
the quadratic form has a zero eigenvalue. For
the CAF state with $E_Z<2|g_{xy}|$, after setting $\cos\theta=\frac{E_Z}{2|g_{xy}|}$,
we find
\begin{widetext}
\beqr
\tcE=4g_0-4g_z-2g_{nz}-\frac{E_Z^2}{|g_{xy}|}+M_{\zeta\zeta}\zeta^2+M_{\gamma\gamma}\gamma^2-\frac{2DE_Z}{|g_{xy}|}\gamma\zeta,\\
M_{\gamma\gamma}=\frac{E_Z^2}{2g_{xy}^2}(|g_{xy}|+(1-r^2)G_0+(1+r^2)g_z),\label{mgg}\\
M_{\zeta\zeta}=\frac{r^2E_Z^2}{|g_{xy}|}+2\big((1-r^2)G_0+(1+r^2)g_z-(2r^2-1)|g_{xy}|\big).\label{mzz}
\eeqr
\end{widetext}
Recalling the definition of $\tD$ [Eq. (\ref{G0Dtilde})]
we can express Eqs. (\ref{mgg}) and (\ref{mzz}) as
\beqr
M_{\gamma\gamma}=&\tD\frac{E_Z^2}{2|g_{xy}|^2},\\
M_{\zeta\zeta}=&2(\tD-2r^2|g_{xy}|)+\frac{r^2E_Z^2}{|g_{xy}|}.
\eeqr
 The critical value $D_{c1}$ for the CAF case is then
\beqr
D_{c1}^{CAF}=&\frac{|g_{xy}|}{E_Z}\sqrt{M_{\gamma\gamma}M_{\zeta\zeta}}\\
=&\sqrt{\tD\bigg(\tD-2r^2|g_{xy}|+\frac{r^2E_Z^2}{2|g_{xy}|}\bigg)}.
\label{Dc1CAF}\eeqr
For the FM state ($E_Z>2|g_{xy}|$), setting $\theta=0$  we obtain
\beqr
&\tcE=4g_0-4g_z-2g_{nz}-4E_Z+4|g_{xy}|\nonumber\\
&-4D\gamma\zeta+M_{\zeta\zeta}\zeta^2+M_{\gamma\gamma}\gamma^2,\\
&M_{\zeta\zeta}=M_{\gamma\gamma}=2\big(2\tD+E_Z-2|g_{xy}|\big).
\eeqr
In this case the critical value is
\beq
D_{c1}^{FM}=\tD+E_Z-2|g_{xy}|.
\label{Dc1FM}\eeq

Now let us turn to $D_{c2}$,  the critical value of $D$ where the KEK/FLP
or the SVE/FLP state is unstable to the three-angle-ansatz.
We start from large $D$ where the FLP
state is clearly the HF ground state.
In this case, since $\theta\approx0$, $\chi\approx0,\ \psi\approx\pi$
we can assume $\theta\ll1$,
$\gamma=\frac{\pi}{2}-\xi,\ \zeta=\frac{\pi}{2}-\omega$, and expand
the energy function for small $\theta,\ \xi,\ \omega$. Due to the fact
that the energy function Eq. (\ref{energy3angle}) depends on $\theta$
only via $\cos\theta$, we see that the quadratic fluctuations of
$\theta$ decouple from those of $\gamma,\ \zeta$. The quadratic
instability of the FLP state in the $\theta$ channel occurs at
\beq
D_{c}^{\theta}=D_c^{FLP/KEK}=g_K
\eeq
and leads to the KEK state which we have already
described.

Ignoring the $\theta$ flucuations, the energy function near the FLP
state can be expanded for small $\xi,\ \omega$ as
\beq
\tcE=\tcE_{FLP}+2(D-4g_z+\tD)(\xi^2+\omega^2)-4E_Z\xi\omega.
\eeq
This leads to
\beq
D^{\xi,\omega}_{c,FLP}=D_c^{FLP/SVE}=4g_z-\tD+E_Z.
\label{Dc2FLP}\eeq
This instability leads to the SVE state which we have also
described. Using Eqs. (\ref{G0Dtilde}) and (\ref{gK}), we note that
\beq
D_c^{FLP/SVE}=D_c^{FLP/KEK}-2r^2|g_{xy}|+E_Z
\label{Dc2SVE_KEK}\eeq
implying that the SVE (KEK) is favored for $E_Z>2r^2|g_{xy}|$ ($E_Z<2r^2|g_{xy}|$).
In either case, the linear instability of the FLP state leads to a
state with a {\em single} broken U(1). Thus, in
order to see where the BU(1)$^2$ state terminates as $D$ increases from $D_{c_1}$,
we need to consider the linear instabilities of the
KEK and SVE states.

First consider the KEK
state, which is stable when $D<g_K$. Once again the $\theta$ fluctuations decouple from
those of the other two angles. The energy function to
quadratic order in $\xi,\ \omega$ is
\beqr
\tcE=&\frac{\cE_{KEK}}{N_\phi}+\frac{4r^2D^2|g_{xy}|}{g_K^2}\xi^2-\frac{4DE_Z}{g_K}\xi\omega\nonumber+M_{\omega\omega}\omega^2,\\
&M_{\omega\omega}= 4r^2|g_{xy}|-2\bigg(1-\frac{D^2}{g_K^2}\bigg)\tD.
\eeqr
We infer the value of $D_{c2}$ from this equation to be
\beq
D_{c2}^{KEK}=g_K\sqrt{1-\frac{2r^2|g_{xy}|}{\tD}+\frac{E_Z^2}{2r^2\tD|g_{xy}|}}.
\label{Dc2KEK}\eeq
In order for the $KEK$ state to be stable we must impose
$D_{c2}<g_K$,  consistent with the requirement
$E_Z<2r^2|g_{xy}|$. We thus identify
a first parameter regime in which BU(1)$^2$ state is the groundstate
for a non-vanishing range of $D$.

For $E_Z>2r^2|g_{xy}|$ the KEK state has no linear
instabilities. If its energy crosses that of the CAF/FM state it must
do so as a first-order transition.

Now we turn to the linear instabilities of the SVE state. The SVE
state corresponds to $\theta=\xi=\pi$ while $\psi$ is nontrivial. The
$\theta$ fluctuations once again decouple from the $\xi,\ \psi$
fluctuations. The $\xi,\ \psi$ fluctuations are innocuous, but the
$\theta$ fluctuations do lead to an instability. A straightforward
analysis shows that
\beq
D_{c2}^{SVE}=D_{min}^{SVE}+\frac{2|g_{xy}|-E_Z}{|g_{xy}|}(g_z-G_0)\; .
\eeq
Recalling the condition for the existence of the SVE state to be
$D_{min}^{SVE}<D<D_{min}^{SVE}+2(1-r^2)(g_z-G_0)$, we indeed see that this is
an actual instability only for $E_Z>2r^2|g_{xy}|$.   We then arrive
at a second scenario in which the BU(1)$^2$ state is stable, in this case
connecting either the CAF state at $D=D_{c_1}^{CAF}$ (for $E_z < 2|g_{xy}|$)
or the FM state at $D=D_{c_1}^{FM}$ (for $E_z > 2|g_{xy}|$) to the SVE state at $D=D_{c_2}^{SVE}$.

Finally, if $E_Z<2r^2|g_{xy}|$ and the energy of the SVE state crosses that of
the CAF/FM, it must do so as a first-order transition.

\section{Main Results and Phase Diagrams}
\label{Results}

As seen in the previous section, there are several different states
that compete in different regimes of $B_{\perp},\ E_Z,\ D$. We will
assume that all the couplings $g_i$ are proportional to
$B_{\perp}$. It would then naively appear that one can scale out
$B_{\perp}$ from the Hamiltonian. However, recall that the parameters
$r$ and $\e_a$ depend on $B_{\perp}$ via their dependence on $\lambda$
[see Eq. (4)] arising from the trigonal warping coefficient
$t_3$.
\begin{figure}
\includegraphics[width=1.0\linewidth]{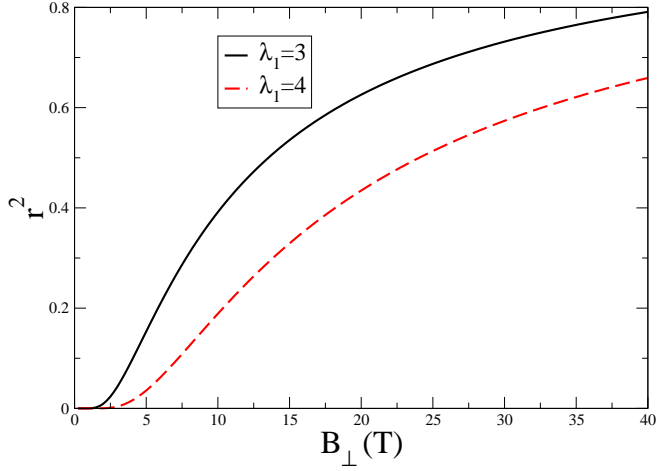}
\caption{$r^2$ vs. $B_\perp$ for $\lambda_1=3$ and
  $\lambda_1=4$. Note that $r^2$ tends vary rapidly to zero when
  $B_\perp$ falls below a characteristic scale set by $\lambda_1$. At
  large values of $B_\perp$, $r^2\to1$. The approach to $r^2=1$ is very
  slow.
\break\break\break}
\label{rsqvsBperp}
\end{figure}

\begin{figure}
\includegraphics[width=1.0\linewidth]{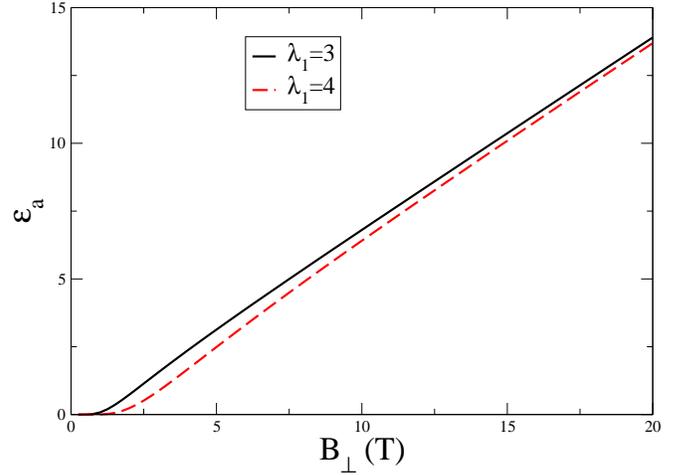}
\caption{Orbital anisotropy energy $\e_a$ vs. $B_\perp$
  for $\lambda_1=3$ and $\lambda_1=4$.  At very small values of $B_\perp$ below a characteristic scale set by $\lambda_1$, $\e_a$ vanishes rapidly as $B_\perp\to 0$. At large values of $B_\perp$, $\e_a$ becomes linear in $B_\perp$. }
\label{anisovsBperp}
\end{figure}

Introducing a field-independent parameter
$\lambda_1=\lambda\sqrt{B_\perp}$ (which is the value of $\lambda$ at
$B_\perp=1$T), in Fig. \ref{rsqvsBperp} and Fig. \ref{anisovsBperp}
we show $r^2$ vs. $B_\perp$ and $\e_a$ vs. $B_\perp$ for $\lambda_1=3$
and 4. We see that both $r^2$ and $\e_a$ vanish very rapidly for
$B_\perp$ smaller than a characteristic scale $B_\lambda$. For
$B_{\perp}\gg B_\lambda$, we see that $r^2\to 1$ while $\e_a$ becomes
linear in $B_\perp$. There are thus two regimes in which the analysis
becomes simple. In the small $B_\perp$ regime we can essentially set
$r^2\approx0$. In the large $B_\perp$ regime we can set
$r^2\approx1$. With the parameters we use the small $B_\perp$ regime
is far easier to realize at experimentally feasible values of
$B_\perp$.

Before presenting the numerical HF results we analyze
the phase diagram for small and large $B_\perp$
analytically. This provides us with relations between the
couplings $g_i$ that determine the topology of the phase diagram.

\subsection{Possible Phase Diagrams at Small $B_\perp$}
\label{smallbperp}

The key idea is to analyse the ordering of the various special values
of $D$ that we defined in Section \ref{instabilities} in the limit
$r^2\to0$. They are
\beqr
&\tD\approx g_z+G_0+|g_{xy}|,\\
&g_K\approx 3g_z-G_0-|g_{xy}|\approx 4g_z-\tD,\\
&D_{c1}^{CAF}\approx \tD,\\
&D_{c1}^{FM}=\tD+E_Z-2|g_{xy}|,\\
&D_{c2}^{SVE}=\tD+E_Z-2|g_{xy}|+\frac{2|g_{xy}-E_Z}{|g_{xy}|}(g_z-G_0).
\eeqr
We have not included $D_{c2}^{KEK}$ because the condition for it to
exist, $E_Z<2r^2|g_{xy}|$, cannot be satisfied when $r^2\to0$. The
condition for the BU(1)$^2$ state to be the true HF
ground state is $D_{c2}>D_{c1}$.  For $E_Z<2|g_{xy}|$, this becomes
\beq
g_z>G_0+|g_{xy}|.
\eeq
If $E_Z>2|g_{xy}|$ then $D_{c2}^{SVE}$ ceases to be physical (because
it becomes less than $D_{min}^{SVE}$). In this case there is no
BU(1)$^2$ state. Instead, as $D$ increases, the FM state gives way to
the SVE state at $D_{c1}^{FM}$, which in turn continuously evolves
to become the FLP state at $D^{FLP/SVE}_{c}=4g_z-\tD+E_Z$, as long as
\beq
g_z>G_0.
\eeq
Thus, we obtain the following three possibilities at small $B_\perp$:
(i) If $g_z<G_0$ there will be a direct first-order transition of the
CAF/FM into either of the SVE/FLP states at all values of $E_Z$. (ii) If
$G_0<g_z<G_0+|g_{xy}|$ then there will be a direct first-order
transition between the CAF and FLP/SVE states as $D$ increases as long
as $E_Z<2|g_{xy}|$. However, for $E_Z>2|g_{xy}|$, the SVE state
smoothly interpolates between the FM at small $D$ to the FLP
state at large $D$. All transitions will now be continuous. (iii) If
$g_z>G_0+|g_{xy}|$, then the BU(1)$^2$ state always
intervenes between the CAF and the SVE states as $D$ is increased for
$E_Z<2|g_{xy}|$. However, for $E_Z>2|g_{xy}|$, the BU(1)$^2$ state
disappears, and instead the SVE smoothly connects the FM and FLP
states.

Now we consider the POP state, and the criteria for whether it is the
true ground state for the small $B_\perp$ regime in which
$r^2\to0$. Some insight can be obtained as follows. Consider
the interlayer potential $D \equiv D_{FLP}^*$ at which the
the CAF and FLP states are equal in energy [Eqs. (\ref{tcecaf}) and (\ref{tceflp})].
Recall that the slope of
the POP state with respect to $D$ is $-2$, while that of the FLP state
is $-4$. We evaluate the energy of the POP state at $D=D_{FLP}^*$. If
$\tcE_{POP}(D_{FLP}^*)>\tcE_{CAF}$ then the POP state will not be the
ground state for any $D$. For purely perpendicular field, assuming
$E_Z\ll2|g_{xy}|$, we have $D_{FLP}^*\approx2g_z$. Since for
$r^2\to0$ we have $\e_a\approx0$, this leads to the
condition for the absence of the POP state,
\beq
G_0+|g_{xy}|-g_z+\frac{3}{2}g_{nz}>0.
\eeq
Recall that in order to see the BU(1)$^2$ state at minimal $E_Z$, and
assuming $E_Z\ll4|g_{xy}|$, we need $g_z>G_0+|g_{xy}|$.  This means in
order for the BU(1)$^2$ state to be the lowest in energy among the
orbitally symmetric states, and for it to have a lower energy than the
POP state, we need $g_{nz}$ greater than some critical value. This is
easily understood, as a large, positive $g_{nz}$ penalizes orbital
polarization.

Let us now turn to the other extreme, very large values of $B_\perp$
such that $r^2\to1$ and $\e_a=\e_{a0}B_\perp$.

\subsection{Possible Phase Diagrams at large $B_\perp$}
\label{largebperp}

Setting $r^2\approx1$ we find
\beqr
\tD&\approx 2g_z+|g_{xy}|,\\
g_K&\approx 2g_z+|g_{xy}|=\tD,\\
D_{c1}^{CAF}&\approx \sqrt{\tD\big(\tD-2|g_{xy}|+\frac{E_Z^2}{2|g_{xy}|}\big)},\\
D_{c1}^{FM}&=\tD+E_Z-2|g_{xy}|=2g_z-|g_{xy}|+E_Z,\\
D_{c2}^{KEK}&\approx\sqrt{\tD\big(\tD-2|g_{xy}|+\frac{E_Z^2}{2|g_{xy}|}\big)}.
\eeqr
For $E_Z<2|g_{xy}|$ we see that $D_c^{FLP/KEK}=g_K>D_{c}^{FLP/SVE}$, which means that
one should consider $D_{c1}^{CAF}$ and $D_{C2}^{KEK}$. However, in the
$r^2\to1$ limit, these are identical! This means the window for the
BU(1)$^2$ state shrinks to zero as $r^2\to1$. The same
is true for $E_Z>2|g_{xy}|$.

At $r^2=1$ and $D=D_{c1}^{CAF}=D_{c2}^{KEK}$ the energy becomes
independent of two of the three angles. This implies a $q=0$
collective mode whose energy vanishes, as has been found in a recent
calculation \cite{denova_2017}. Thus, hints of the potential existence
of the BU(1)$^2$ state can be seen in the collective mode spectrum
even at $r^2=1$.

We see then that the trigonal warping $t_3$, via the parameter
$r^2<1$, is responsible for the existence of the BU(1)$^2$ state in a
nonvanishing region of the parameter space. For this reason previous
theoretical analyses, which in general have not included the effects
of $t_3$, have not identified this state in the phase diagram.

\subsection{Hartree-Fock Phase Diagrams}
\label{hfphasedias}

Since the space of couplings is so large, we will take some guidance
from experiments to narrow our choices. The POP state has been seen in
experiments on BLG at $\nu=0$: at purely perpendicular fields, it
makes its appearance for $B_\perp>$12 T \cite{Hunt_2016}. In some
experiments a direct transition\cite{Hunt_2016} is seen between a
putative CAF state at small $D$ and a putative FLP state at larger
$D$, while in others there are intriguing hints that there may be an
intermediate phase between the CAF and the FLP at small $B_\perp$
\cite{maher_2013,jun-pvt}.  Presumably, disorder, the screening
environment, or perhaps microscopic features of how the samples are
prepared, determine whether the intermediate phase is seen. A second
result we will take from experiments is that when one tries to fit the
observed sequence of transitions to a single-particle model, the
anisotropy energy appears to be close to zero for $B_\perp<10$ T but
turns on afterwards \cite{li_2017}. Looking at Fig. \ref{anisovsBperp}
we see that there is a similar behavior of $\e_a$ vs. $B_\perp$. This
allows us to conjecture that the effective value of $\lambda$ is
rather larger than conventionally assumed.

To account for this diversity of observations, we will consider three
sets of parameters embodying the three regimes of $g_z$ that we
obtained in Section \ref{smallbperp} for small $B_\perp$. {\bf Parameter Set 1} (PS1) will have
$g_z>G_0+|g_{xy}|$, so that there is an intervening Broken-U(1)$\times$U(1) phase as a function of $D$ between the CAF and the FLP phases
for $E_Z<2|g_{xy}|$. {\bf Parameter Set 2} (PS2) will have
$G_0<g_z<G_0+|g_{xy}|$. This means that at the
minimal $E_Z$ there is a direct first-order transition between the CAF
and SVE phases, while for $E_Z>2|g_{xy}|$ the SVE phase smoothly connects the FM state to the FLP state. {\bf Parameter Set 3} (PS3) will
have $G_0>g_z$, so that there is always a direct first-order
transition between the CAF and FLP phases.

\subsubsection{Parameter Set 1}
\label{PS1}

The values we use (arbitrary units) are $g_0=0.5B_\perp$,
$g_z=3.5B_\perp$, $g_{xy}=-1.65B_\perp$, and $g_{nz}=1.0B_\perp$.
The dimensionless parameter
$\lambda$ of Eq. (\ref{lambda_def}) is assumed to be
$\lambda=5.0/\sqrt{B_\perp}$. In order to keep the POP state from appearing below about $12$T, we set the orbital anisotropy to $\tea=1.4$

Since we are using arbitrary units for the couplings $g_i$, our
results for the values of $D$ at which transitions take place are also
arbitrary. Therefore, in the phase diagrams that follow, we will not
put units on the $D$ axis.

\begin{figure}
\includegraphics[width=1.0\linewidth]{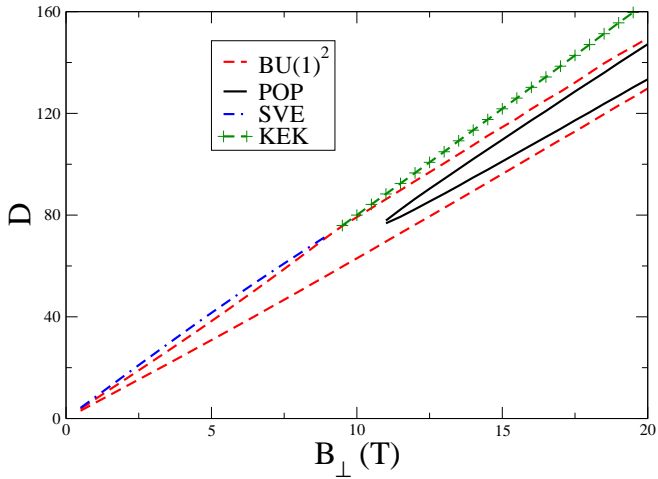}
\caption{The $B_\perp-D$ phase diagram for PS1 for the case of only
  perpendicular field. Here and in the following, $D$ is in arbitrary
  units. This is identical to Fig. 1, reproduced here for convenience.
  At small $D$, the system is always in the CAF phase. For
  $B_\perp<11$T, as $D$ increases, the system undergoes a second-order
  phase transition (dashed red line) to the BU(1)$^2$ phase. Another
  second-order phase transition (dashed red line) takes the system at
  a slightly higher $D$ to the SVE state. Finally, at an even higher
  $D$ (dash-dotted blue line) the system goes into the FLP phase. For
  $B_\perp>11$T the POP state becomes lower in energy than the
  BU(1)$^2$ state for an intermediate range of $D$, and is the ground
  state between the two solid black lines.  At higher $B_\perp$ the
  BU(1)$^2$ state gives way to the KEK state at the dashed red line,
  which in turn gives way to the FLP state at the green dashed line
  with the + symbols. }
\label{PS1-bperp-phase-dia-fig2}
\end{figure}

\begin{figure}
\includegraphics[width=1.0\linewidth]{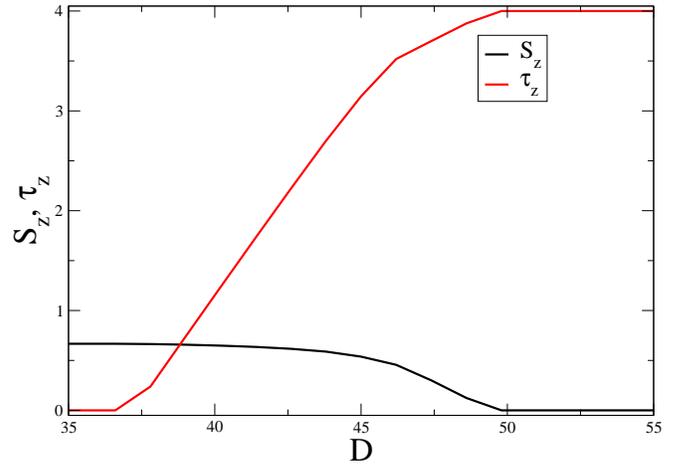}
\caption{Order parameters at $B_\perp=6$T for PS1 with purely
  perpendicular field.  Recall that $D$ is in arbitrary
  units. For $D<36$, the system is in the CAF phase. It makes a
  second-order transition to the BU(1)$^2$ phase at $D=36$ and remains
  in this phase till $D=46$, at which point it makes another
  second-order transition into the SVE phase. The SVE phase persists
  till about $D=50$, beyond which the system is fully layer
  polarized. \break}
\label{Order-para-PS1-bperp-cut-6p0-fig}
\end{figure}

Let us first consider the case of a perpendicular field only. From
experimental measurements \cite{jun-pvt}, the total field needed
to spin-polarize a sample at $B_\perp=2$T is about $12$T. We
combine this with the theoretical critical Zeeman coupling for full
spin-polarization, $E_Z=2|g_{xy}|$, to obtain
$E_Z=\frac{1}{3}|g_{xy}|$ for a purely perpendicular field. The phase
diagram for this situation is shown in Fig.  \ref{PS1-bperp-phase-dia-fig2}.
As can be seen, most of the phases discussed
before appear in the phase diagram. Let us first focus on the small
$B_\perp$ region, where we expect $r^2\ll1$. In
accordance with the expectations of Section \ref{smallbperp}, we see
that with increasing $D$, one encounters, in order, the
CAF, BU(1)$^2$, SVE, and FLP states, all of which are identified
from the numerically generated $\Delta$ matrix.
Fig. \ref{Order-para-PS1-bperp-cut-6p0-fig} illustrates the spin polarization
$S_z$ and the valley polarization $\tau_z$ at fixed
$B_\perp=6$T as a function of $D$.
At this field
the CAF gives way to the BU(1)$^2$ state at around $D=36$. At
$D\approx46$ a slight kink in the lines indicates that the system has
made a transition to the SVE state.  The SVE state is stable in the interval
$46\leqslant D\leqslant 50$, and for $D>50$ the system is in
the FLP state.

\begin{figure}
\vskip 0.5cm
\includegraphics[width=1.0\linewidth]{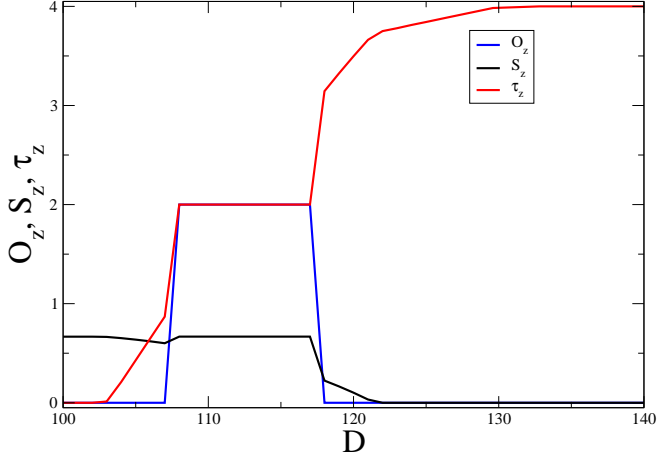}
\caption{Order parameters at $B_\perp=16$T for PS1 with purely perpendicular
field. Once again, the system is in the CAF
  state for small $D$. At around $D=102$, there is a second-order
  transition into the BU(1)$^2$ state. This is followed by a
  first-order transition into the POP state at $D=107$.  The POP state
  persists until $D=118$, at which point the system makes a
  first-order transition back into the BU(1)$^2$ state. At about
  $D=122$ there is another second-order transition, this time into the
  KEK state. Finally, at abour $D=130$, the KEK state gives way to the
  FLP state. \break}
\label{Order-para-PS1-bperp-cut-16p0-fig}
\end{figure}

At larger $B_\perp>11$T, the POP state makes its appearance by
``eating-up'' some of the regime that belongs to the BU(1)$^2$ state. An illustrative cut at $B_\perp=16$T is shown in Fig. \ref{Order-para-PS1-bperp-cut-16p0-fig}, which in
addition to $S_z$ and $\tau_z$ illustrates $O_z$, the orbital polarization.
Now we see that the system undergoes a second-order transition from
the CAF state to the BU(1)$^2$ state at $D\approx102$. This is
followed by a first-order transition to the POP state at
$D\approx107$, which then persists until $D\approx118$. The system now
undergoes a first-order transition to a narrow sliver of the BU(1)$^2$
state, which gives way to the KEK state at $D\approx122$. The KEK
state persists until $D\approx130$ beyond which the system is in the
FLP state.

\begin{figure}
\includegraphics[width=1.0\linewidth]{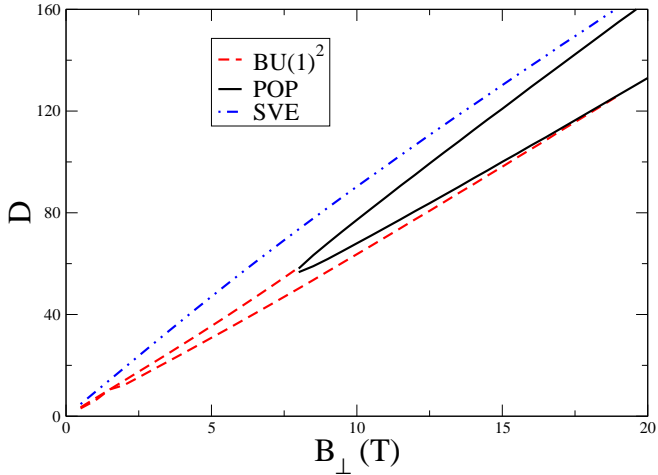}
\caption{Phase diagram for PS1 in a tilted field, such that
  $E_Z=|g_{xy}|$. The BU(1)$^2$ state appears between the dashed red
  lines, while the POP state appears between the solid black lines. The
  main qualitative difference between this figure and
  Fig. \ref{PS1-bperp-phase-dia-fig2} is the absence of the KEK state at large
  $B_\perp$, where it has been supplanted by the SVE state.  All
  transitions are second-order except for those into and out of the
  POP state. \break\break}
\label{PS1-EZeqgxy-phase-dia-fig}
\end{figure}

For completeness, we present two other phase diagrams. In
Fig. \ref{PS1-EZeqgxy-phase-dia-fig}, we consider an intermediate
value of tilted field with $E_Z=|g_{xy}|$. The low $D$ phase is still
the CAF state.
Note that the SVE state expands its domain compared to perperdicular
field, and the BU(1)$^2$ state has a correspondingly smaller
domain. The KEK state has disappeared altogether. This is because,
unlike the SVE state, it has no spin polarization and thus cannot take
advantage of the Zeeman field. The domain of the POP state has also
expanded, and now it reaches down to $B_\perp=8$T.

\begin{figure}
\includegraphics[width=1.0\linewidth]{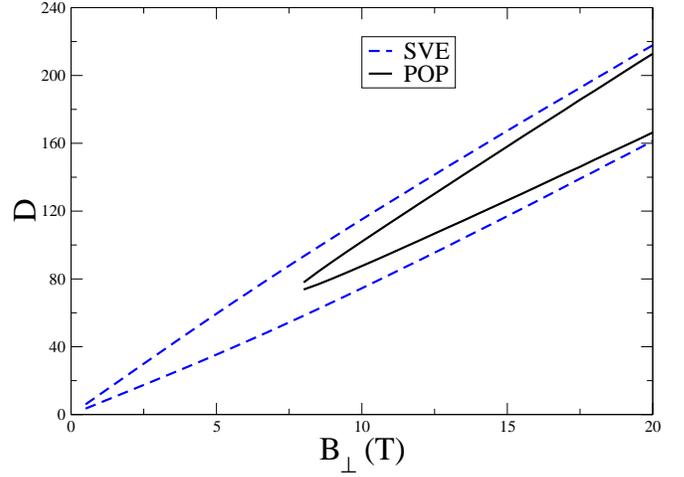}
\caption{Phase diagram for PS1 at $E_Z=2.5|g_{xy}|$. The small $D$
  region is now in the fully spin-polarized FM state, which makes a
  second-order transition (lower dashed blue line) to the SVE state,
  which in turn gives way to the FLP state via another second-order
  transition (upper dashed blue line). The BU(1)$^2$ state has
  disappeared and has been supplanted by the SVE state. The POP state intrudes into the SVE region via first-order transitions (solid black lines). \break\break }
\label{PS1-EZ2p5gxy-phase-dia-fig}
\end{figure}

In Fig. \ref{PS1-EZ2p5gxy-phase-dia-fig}, we present the phase diagram
for a very large tilted field of $E_Z=2.5|g_{xy}|$. The low $D$ phase
is now the FM state.
We see that the BU(1)$^2$ state has disappeared. The SVE and POP
states are better able to take advantage of the large $E_Z$ at
intermediate values of $D$.

\subsubsection{Parameter Set 2}
\label{PS2}
\begin{figure}
\includegraphics[width=1.0\linewidth]{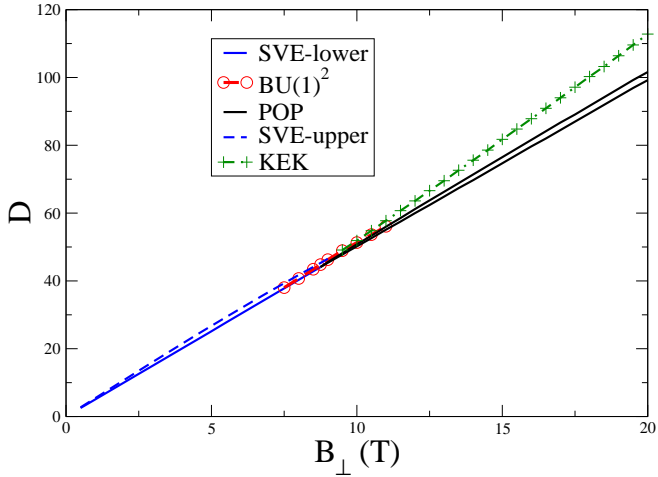}
\caption{Phase diagram for PS2 at $E_Z=|g_{xy}|/3$ (purely
  perpendicular field.) The small $D$ region is in the CAF phase. For
  small $B_\perp<8$T the CAF makes a direct first-order transition
  into the SVE phase (solid blue line), which then gives way to the
  FLP phase via a second-order transition (dashed blue line). Between
  8T and 10T, the situation is very complicated at intermediate $D$,
  where many phases are almost identical in energy. At 10T, as one
  increases $D$, there is a direct first-order transition from the CAF
  phase into the POP state (lower solid black line). The system exits
  the POP state via another first-order transition into a narrow
  sliver of the BU(1)$^2$ state, which exists between the upper solid
  black line and the red line with circles. The BU(1)$^2$ state enters
  the KEK state via a second-order phase transition. Finally, the KEK
  state gives way to the FLP state. At larger $B_\perp$ the situation
  simplifies: The CAF makes a first-order transition into the POP,
  which makes another first-order transition into the KEK, which
  finally makes a second-order transition to th FLP state (dashed
  green line with + symbols).\break\break}
\label{PS2-bperp-phase-dia-fig}
\end{figure}

This set of parameters is identical to PS1, except $g_z=2.5B_\perp$. This change means that now
$G_0<g_z<G_0+|g_{xy}|$. Furthermore, to keep the POP state from appearing below $\simeq 10$T, we need to increase the dimensionless orbital anisotropy to $\tea=1.77$.
Fig. \ref{PS2-bperp-phase-dia-fig} shows the
phase diagram for PS2 with a purely perpendicular field
($E_Z=|g_{xy}|/3$). As can be seen, the BU(1)$^2$ phase has
almost disappeared from the phase diagram. There is a tiny remnant of
it for 8T$<B_\perp<$10T.

\begin{figure}
\includegraphics[width=1.0\linewidth]{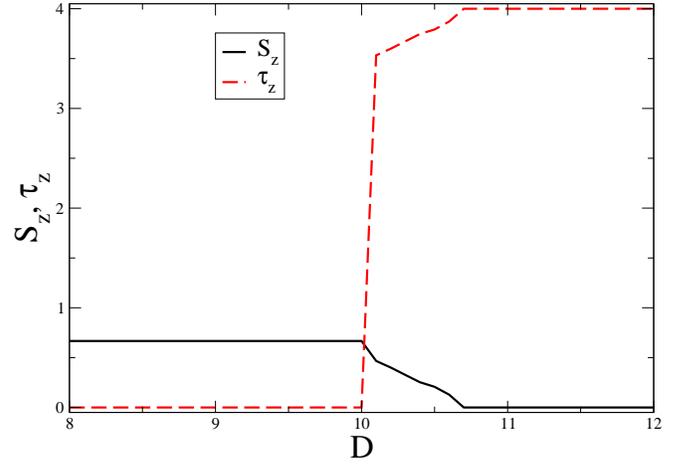}
\caption{Order parameters as a function of $D$ at $B_\perp=2$T in PS2
  for purely perpendicular field.  The first-order nature of the
  transition between the CAF and the SVE states is clear. The SVE
  order parameters smoothly go over to those of the FLP. \break\break}
\label{Order-para-PS2-bperp-cut-2p0-fig}
\end{figure}

\begin{figure}
\includegraphics[width=1.0\linewidth]{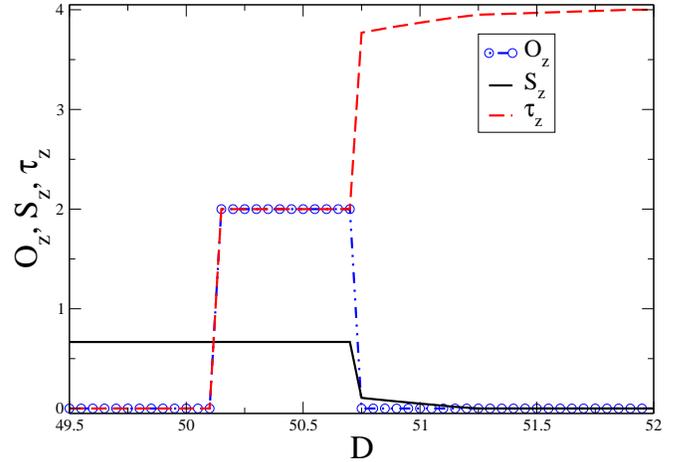}
\caption{Order parameters as a function of $D$ at
  $B_\perp=10$T in PS2 for purely perpendicular field. At small $D$ the system is in the CAF
  phase. It makes a first-order transition into the POP state at
  $D=50.1$. The POP state gives way to the BU(1)$^2$ state via a
  first-order transition at $D=50.75$.  The BU(1)$^2$ state persists
  until $D=51.25$, at which point the system makes a second-order
  transition to the KEK state.  Finally, at $D=52$, the KEK state gives way to the FLP state via a second-order transition.
\break\break}
\label{Order-para-PS2-bperp-cut-10p0-fig}
\end{figure}

\begin{figure}
\includegraphics[width=1.0\linewidth]{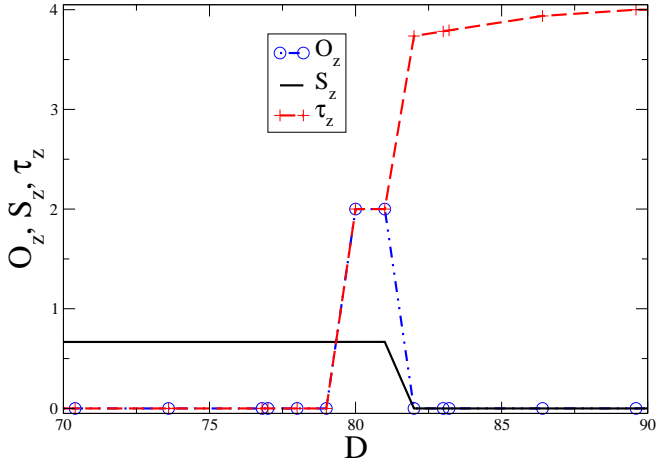}
\caption{Order parameters as a function of $D$ at
  $B_\perp=16$T in PS2 for purely perpendicular field. As $D$ increases, the first two transitions,
  from the CAF into the POP, and from the POP into the KEK state, are
  first-order. The final transition from the KEK to the FLP state is
  second-order. \break\break}
\label{Order-para-PS2-bperp-cut-16p0-fig}
\end{figure}

There are several differences in the phase diagrams between PS1 and
PS2. Focusing first on small $B_\perp$, the CAF goes into the SVE
phase via a first-order transition, without going through the
BU(1)$^2$ phase. The SVE phase gives way to the FLP phase at larger
$D$ via a second-order
transition. Fig. \ref{Order-para-PS2-bperp-cut-2p0-fig} shows the
evolution of the order parameters with $D$ for fixed $B_\perp=2$T.
In Fig. \ref{Order-para-PS2-bperp-cut-10p0-fig} we show the evolution
of the order parameters at $B_\perp=10$T, which includes a sliver of
the BU(1)$^2$ state.
The evolution of the order parameters at $B_\perp=16$T is presented in Fig. \ref{Order-para-PS2-bperp-cut-16p0-fig}.

\begin{figure}
\includegraphics[width=1.0\linewidth]{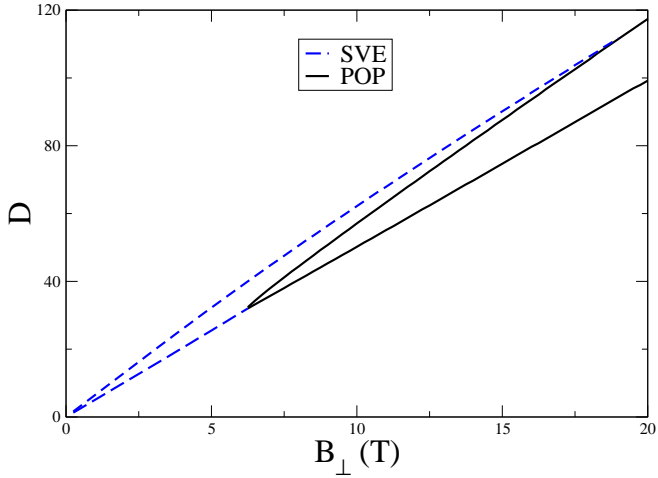}
\caption{Phase diagram for PS2 in a tilted field, such that
  $E_Z=|g_{xy}|$. Only the CAF, the SVE, FLP and the POP appear. The
  transitions between the CAF, SVE and FLP states (dashed blue lines)
  are second-order, while those from the POP state (solid black lines) are
  first-order. \break\break}
\label{PS2-EZeqgxy-phase-dia-fig}
\end{figure}

For completeness we examine PS2 for larger Zeeman values. In
Fig. \ref{PS2-EZeqgxy-phase-dia-fig} we present the phase diagram for
PS2 at $E_Z=|g_{xy}|$. For $B_\perp<8$T, there are only two
transitions as $D$ increases. First the CAF goes into the SVE state
via a first-order phase transition, and then the SVE state gives way
to the FLP state via a second-order transition. For larger
$B_\perp>8$T, the CAF goes directly into the POP state via a
first-order transition. The system then makes another first-order
transition into the SVE state, which finally undergoes a second-order
transition into the FLP state. Note also that the POP state, being
able to take advantage of the larger Zeeman coupling, now appears at
smaller values of $B_\perp$ as compared to the case of perpendicular
field only.
%

\begin{figure}
\includegraphics[width=1.0\linewidth]{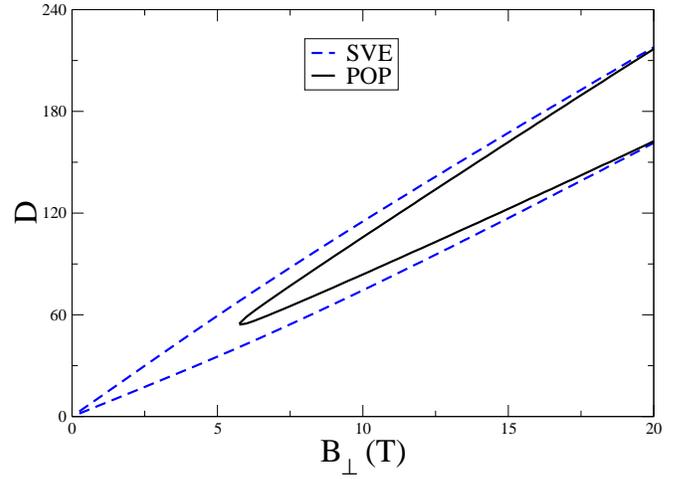}
\caption{The phase diagram for PS2 in a large Zeeman field $E_Z=2.5|g_{xy}|$. Only the FM phase at small $D$, the SVE, the FLP and the POP phases appear. \break\break}
\label{PS2-EZ2p5gxy-phase-dia-fig}
\end{figure}

\begin{figure}
\includegraphics[width=1.0\linewidth]{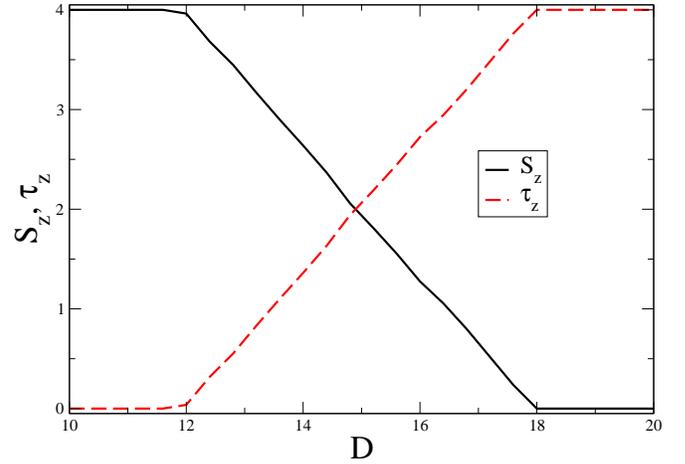}
\caption{Order parameters for $B_\perp=2$T in PS2 at large Zeeman coupling, such that $E_Z=2.5|g_{xy}|$. The small $D$ phase is the fully spin-polarized FM. This makes a second-order phase transition into the SVE, which smoothly interpolates to the FLP state via another second-order phase transition. \break\break}
\label{Order-para-PS2-EZ2p5gxy-cut-2p0-fig}
\end{figure}

\begin{figure}
\includegraphics[width=1.0\linewidth]{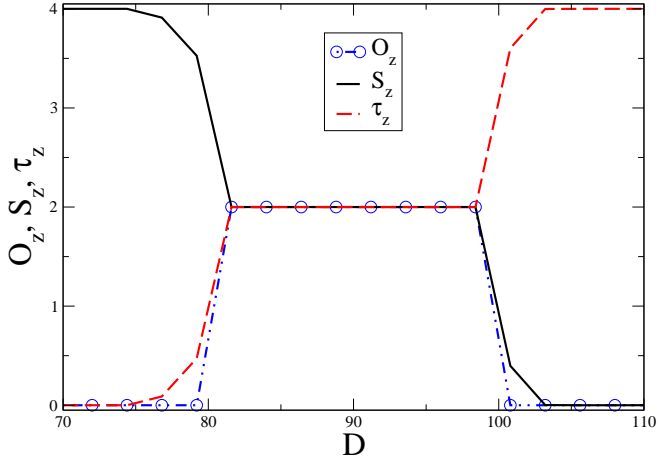}
\caption{Order parameters at $B_\perp=12$T in PS2 at a large Zeeman coupling $E_Z=2.5|g_{xy}|$. The
  small $D$ phase is the fully spin-polarized FM. This makes a
  second-order phase transition into the SVE. The POP state intrudes
  via a first-order transition into the SVE. Another first-order
  transition takes the system back into the SVE, which smoothly
  interpolates to the FLP state via another second-order phase
  transition. \break\break\break\break}
\label{Order-para-PS2-EZ2p5gxy-cut-12p0-fig}
\end{figure}

In Fig. \ref{PS2-EZ2p5gxy-phase-dia-fig} we present the phase diagram for PS2 at large Zeeman coupling, $E_Z=2.5|g_{xy}|$.
The low $D$ phase is now the FM state. This implies that the
transition from the FM to the SVE state should be second-order, since the
SVE smoothly interpolates between the FM and the FLP. Indeed, in
Fig. \ref{Order-para-PS2-EZ2p5gxy-cut-2p0-fig}, a cut at $B_\perp=2$T
showing the evolution of the order parameters as a function of $D$
exhibits the second-order nature.
At larger values of $B_\perp$, the POP state intervenes and two
additional first-order phase transitions, into and out of the POP
state, appear, as seen in
Fig. \ref{Order-para-PS2-EZ2p5gxy-cut-12p0-fig}.

\subsubsection{Parameter Set 3}
\label{PS3}

For PS3, we need to have $g_z<G_0$. So we choose the following values:
$g_0=1.5B_\perp,\ g_z=1.75B_\perp,\ g_{xy}=-1.65B_\perp,\ g_{nz}=B_\perp$, and keep
$\lambda_1=5$. In order to have the POP state not appear below
$B_\perp=12$T at purely perpendicular field, we have to increase the
value of the dimensionless orbital anisotropy to $\tea=3.8$.

\begin{figure}
\includegraphics[width=1.0\linewidth]{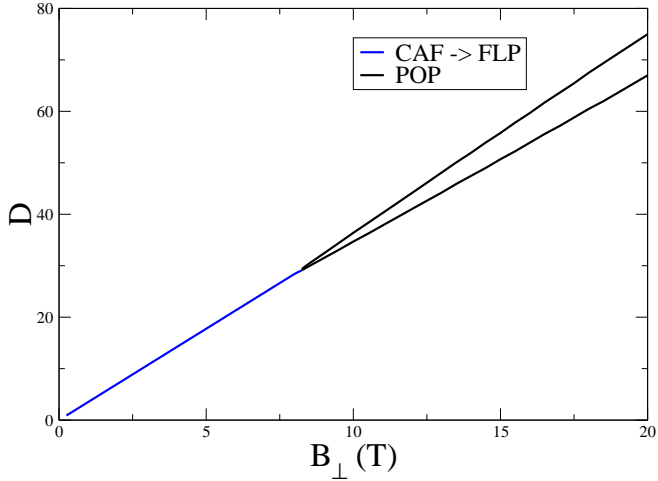}
\caption{Phase diagram for PS3 at purely perpendicular field, $E_Z=|g_{xy}|/3$. All the transitions are first-order. The CAF gives way directly to the FLP at small $B_\perp$, whereas the POP state intrudes for larger $B_\perp$. \break\break}
\label{PS3-bperp-phase-dia-fig}
\end{figure}

In Fig. \ref{PS3-bperp-phase-dia-fig} we show the phase diagram for
PS3 at purely perpendicular field. This is the simplest topology of
the phase diagram, and only the CAF, FLP and POP states appear. All
the transitions are first-order.

\begin{figure}
\includegraphics[width=1.0\linewidth]{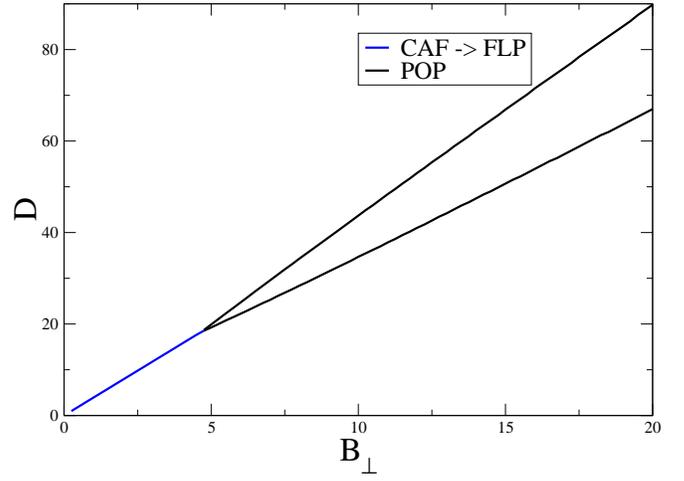}
\caption{Phase diagram for PS3 at an intermediate value of Zeeman coupling, $E_Z=|g_{xy}|$.  This is very similar to the phase diagram of PS3 at perpendicular field. All transitions are first order. \break\break }
\label{PS3-EZeqgxy-phase-dia-fig}
\end{figure}

In Fig. \ref{PS3-EZeqgxy-phase-dia-fig} we show the phase diagram at
an intermediate value of the Zeeman coupling, $E_Z=|g_{xy}|$. Apart
from the POP state appearing at lower $B_\perp$, and extending to
larger $D$, there are no qualitative differences from the case of
purely perpendicular field.

\begin{figure}
\includegraphics[width=1.0\linewidth]{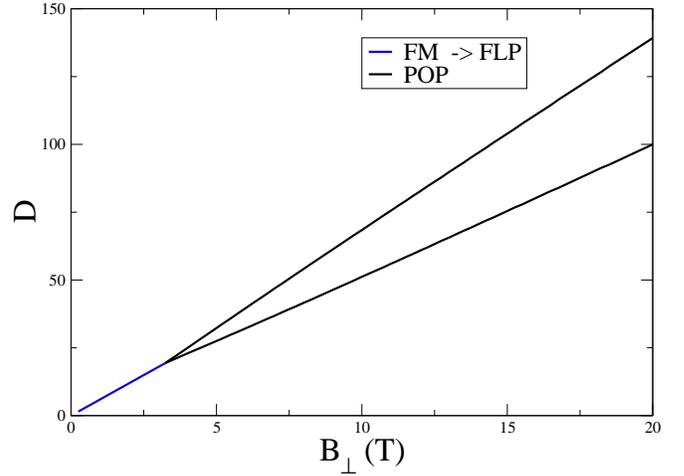}
\caption{Phase diagram for PS3 at a large value of Zeeman coupling, $E_Z=|g_{xy}|$. The small $D$ phase is the FM, otherwise the phase diagram is very similar to those for PS3 at smaller $E_Z$. \break\break}
\label{PS3-EZ2p5gxy-phase-dia-fig}
\end{figure}

\section{Discussion}
\label{discussion}
\subsection{Experimental Signatures of the Phase Transitions}
\label{expt-sig-PT}

We begin this section by discussing possible experimental signatures
of the phases and transitions discussed above.

To our knowledge three types of measurements have been performed on
BLG in the quantum Hall regime: transport, compressibility, and
layer polarizability.  With respect to transport, all the bulk
states we have analyzed are insulators with a charge gap. Deep within
a phase, transport occurs only at the edges. In BLG, all quantum
numbers except spin are broken by the edge potential; because of this,
the FM state is expected to be a quantum spin Hall state
\cite{Abanin_2006,Fertig2006,SFP,pasha} whereas the others are trivial
non-conducting states
\cite{Kharitonov_edge,us_2014,us_2016,kharitonov_2016}. At a
transition between two bulk phases, there can be conduction by two
distinct mechanisms. Firstly, if the transition is second-order and
has at least one broken U(1) symmetry on at least one side of the
transition (all our second-order transitions have this property), we
may expect the stiffness of the broken U(1) angle to vanish at the
transition. This leads to gapless charged {\it edge} excitations, as
the present authors have established in monolayer graphene
\cite{us_2014,us_2016}. Secondly, if the transition is first-order,
one may expect the formation of domains due to disorder. Presumably
charged excitations are attracted to the domain walls, and if they
percolate, there may be {\it bulk} conduction
\cite{jungwirth_2001,dhochak_2015}. Thus, both first- and second-order
transitions are expected to be visible in transport.

Bulk excitations can also provide information about the nature of the
ground state. For example, gapless modes associated with broken U(1)
symmetries should have clear signatures in heat
transport\cite{pientka_2017}. Bulk excitations can also be probed via
the compressibility.  Several of the transitions we have described
involve a U(1) symmetry breaking as the transition is crossed.  In the
broken symmetry phase, near the transition where there is a soft
stiffness one expects very low energy, charged merons
\cite{QHFM}. Nevertheless, we expect the system to remain
incompressible at zero temperature: in order to inject an electron,
one has to combine this low-energy meron (which is expected to support
a small charge) with a high-energy antimeron (carrying the remaining
charge of the electron).  The resulting bimeron, the form in which
electrons can be injected into the system, will have non-vanishing
energy in spite of the low energy of one of its components.  By
contrast, at a first-order transition, if the domain walls percolate
we expect that electrons can be injected at arbitrarily low energy,
and the system becomes compressible. At $T>0$, the key criterion is
whether the phase with spontaneously broken U(1) is below its
Kosterlitz-Thouless transition temperature $T_{KT}$. In particular,
the appearance of unbound (charged) vortices above $T_{KT}$ may lead
to singular behavior in the compressibility as a function of
temperature.

Finally, layer polarizability measurements have recently become
feasible for this system \cite{Hunt_2016}.  The level of charge in
each layer continuously varies in any state for which there is a
broken U(1) symmetry involving the valley degree of freedom.  Thus,
the FM, CAF, POP and FLP states have a vanishing linear layer
polarizability, while the BU(1)$^2$, SVE, and KEK states are layer
polarizable.  Such experiments thus allow one to probe when the
U(1)$_{valley}$ symmetry is spontaneously broken in the bulk.

Current experiments on BLG suggest that the CAF, FM, POP, and FLP
states can be stable in BLG. In a subset of samples, at small
$B_\perp$, an intermediate state\cite{maher_2013,jun-pvt} {\it may}
have been seen between the CAF and the FLP phases, suggesting that
such samples are in parameter regimes consistent with PS1 or PS2. In
some samples, an intermediate phase is also seen at small $B_\perp$,
albeit at large tilted field, between the FM and the FLP
phases\cite{jun-pvt}. Again, this is consistent with both PS1 and
PS2. In other experiments, however, no intermediate phases are seen
between the CAF and the FLP at small $B_\perp$, suggesting that those
samples are consistent with PS3.  What precisely determines in which
parameter regime a particular sample might be remains unclear at
this time, and is a subject for further investigation.  Detailed
observations at small $B_{\perp}$ in extremely clean and cold samples
would greatly clarify the parameter regime to which pure BLG belongs.

It is interesting to carry out a thought experiment in which we assume
that the bulk spin susceptibility $\frac{\partial S_z}{\partial E_Z}$
can be measured, in addition to the layer polarizability
$\frac{\partial \tau_z}{\partial D}$ and the cross-susceptibilities
$\frac{\partial S_z}{\partial D}\equiv \frac{\partial \tau_z}{\partial
  E_Z}$ (the last identity is a Maxwell relation arising from
$S_z=-\frac{\partial \tcE}{\partial E_z}$ and $\tau_z=-\frac{\partial
  \tcE}{\partial D}$). Such measurements may indeed be
  accessible, e.g. using the technique of Reznikov {\it et al}
\cite{Reznikov}. The combined measurement allows one to distinguish
between the different possible states. The FM, CAF, POP, and FLP have
a vanishing layer polarizability. The FM, KEK, and FLP have a
vanishing spin susceptibility. The SVE state has both layer
polarizability and spin susceptibility nonvanishing, but satisfies
$S_z+\tau_z=4$, which implies
\beq \frac{\partial S_z}{\partial D}+\frac{\partial \tau_z}{\partial
  D}=0=\frac{\partial S_z}{\partial E_Z}+\frac{\partial
  \tau_z}{\partial E_Z}
\label{SVE-sus-constraint}\eeq
Finally, the BU(1)$^2$ state also has all susceptibilities
nonvanishing, but is not subject to the condition of
Eq. (\ref{SVE-sus-constraint}). This allows us, in principle at least,
to distinguish the BU(1)$^2$ state from other possibilities.

\subsection{Caveats and Omissions}
\label{Caveats}

We next briefly review some of the underlying assumptions that lead to
the model analyzed in this work. We first separated the Coulomb and
other lattice scale interactions into an SU(4) symmetric part (which
plays no role in choosing the ground state) and a part that does not
respect SU(4) symmetry,. We assumed that the part that does not
respect SU(4) symmetry can be represented as short-range
interactions. These short-range interactions respect the spin-SU(2)
but have only a U(1) symmetry in the valley indices. Finally we
assumed that all interaction parameters $g_i$ are proportional to
$B_\perp$, corresponding to ultra-short-range interactions.

Each of these assumptions can be challenged. Consider first our assumption that
$g_i\propto B_\perp$.  This seems reasonable from the renormalization
group (RG) standpoint, as can be seen from the following argument. At
high energies, the dispersion is Dirac-like, and short-range
interactions are irrelevant as one scales down in energy:
\beq
\tg_i(\ell)=\tg_i(0)e^{-\ell},
\eeq
where $\tg_i$ are the dimensionless couplings (the ratio of the
dimensionful couplings to the kinetic energy scale), $\ell$ is the RG
flow parameter defined by $e^{-\ell}=\Lambda(\ell)/\Lambda(0)$, and
$\Lambda(0)$ is the bandwidth. At a scale proportional to the interlayer
hopping $t_\perp$ (corresponding to RG scale $\ell_\perp$, say) the
quadratic band touching manifests itself, and the one-loop RG flow of
$\tg_i$, if one neglects $t_3$, becomes marginal \cite{Vafek_2010}.
In general the RG flows may be written in the form
\beq
\frac{d\tg_i}{d\ell}=C_{ijk}\tg_j\tg_k,
\eeq
and should be stopped at a kinetic energy scale $\sim B_{\perp}$ which is of relevance
to the system we are studying. Since they are marginal, the values of $g_i$
will follow the kinetic energy scale, thus becoming proportional to
$B_\perp$.

Complications arise when $t_3$ enters the picture. At the
quadratic band touching $t_3$ is a relevant coupling and will
grow. Further, we know that $t_3$ is generated by the
interactions\cite{Pujari_2016}, and will in turn affect the flow of
the $g_i$. Thus, it is likely that the couplings $g_i$ do have some
$B_\perp$ dependence in the presence of trigonal warping. Since we
have not worked out the RG flow equations in the presence of $t_3$, we
have not taken this into account, and have made the na\"ive
assumption that $g_i\propto B_\perp$, which follows from directly
computing the interaction matrix elements for our model
in the Landau levels of interest, without including any renormalization
effects.

Secondly, we assume that all our interactions are
ultra-short-range. Here we are on somewhat firmer footing. Introducing
a $\bq$-dependence of the form $e^{-|\bq|^2\xi^2}$ into the interactions
$v_i(\bq)$ will leave the Hartree terms unchanged, but reduce the
exchange terms by a factor close to unity. This does change some of
the inequalities which we use to define the different parameter sets
(PS1, PS2, and PS3), but does not change the qualitative nature of the
phases or the topologies of the phase diagrams.
As an aside, introducing
such a $\bq$-dependence into the Kharitonov model \cite{kharitonov_bulk_bilayer} will lead to a
BU(1)$^2$ phase in the phase diagram.

Thirdly, we reiterate that the four couplings retained in our
interaction model are only a subset of many such couplings which are
allowed by the symmetry of the system. This was largely to keep a tractable
parameter space size for our study; however, we believe that other couplings will not
qualitatively alter the topologies of the phase diagrams or the
nature of the phases we encounter.

Finally, our analysis has been carried out within the Hartree-Fock approximation. Quantum
fluctuations could play an important role near second-order phase
transitions, particularly for states with broken U(1) symmetries. These are generically
accompanied by soft stiffnesses when they are first entered, so that low-energy excitations
around the HF state will necessarily exist.

\section{Conclusions and Open Questions}
\label{Conclusions}

In this work we have studied the possible zero-temperature ground
states of bilayer graphene (BLG) at charge neutrality in a quantizing
perpendicular magnetic field $B_\perp$.  This $\nu=0$ system is very
rich, possessing three sets of discrete labels: spin, valley, and
orbital, leading to eight nearly degenerate Landau levels in the
low-energy manifold. (Recall that by ``low-energy manifold'' we mean
the manifold of states near the Fermi energy.)
Experimentally, the system can be probed by applying a tilted magnetic
field (to increase the Zeeman coupling $E_Z$) and/or by applying a
perpendicular electric field $D$ which induces layer polarization. In
the presence of these external fields, the symmetry of the problem is
reduced to U(1)$_{spin}\times$U(1)$_{valley}$.

Our philosophy is to ignore the SU(4) symmetric, long-range part of
the Coulomb interaction completely, because it plays no role in ground
state selection at $\nu=0$.  Our model is based on an effective
Hamiltonian, containing only short-range interactions, in the truncated
Hilbert space of the low-energy manifold. Effects of the filled Dirac
sea \cite{Herbut_2007,Shizuya_2012,Roy_2014_1,Roy_2014} are assumed to be
absorbed into renormalizations of the couplings of the effective
Hamiltonian \cite{kharitonov_bulk_monolayer,kharitonov_bulk_bilayer}.

We incorporate two aspects distinct from previous work
\cite{kharitonov_bulk_bilayer,lukose_2016,knothe_2016,Jia_2017}: (i)
We include the effect of the trigonal warping $t_3$ (an interlayer
hopping term allowed by the lattice symmetries) nonperturbatively in
the one-body states of the low-energy manifold that form our
basis. (ii) In addition to interactions introduced in previous work
\cite{kharitonov_bulk_monolayer,kharitonov_bulk_bilayer} ($g_z$ and
$g_{xy}$ which correspond to $U(1)_{valley}$ symmetric interactions),
we introduce two new interactions into our effective Hamiltonian, one
($g_0$) which treats all discrete labels equally, and another
($g_{nz}$) which is an Ising-like interaction in the orbital sector.

The dependence of the dimensionless coupling constant associated with
$t_3$ on $B_\perp$, together with suitable values of the interaction
strengths, leads to the stabilizition of a hitherto unknown
phase. This phase, which we dub the Broken-U(1)$\times$U(1) or
BU(1)$^2$ phase, spontaneously breaks two distinct U(1) symmetries,
and is one of the central findings in this work. Hints of its
existence can be gleaned from unexpected zero modes in the collective
spectrum \cite{denova_2017} even at $t_3=0$. In contrast, all phases
known previously at $\nu=0$ are either symmetric under
U(1)$_{spin}\times$U(1)$_{valley}$ or spontaneously break a single
U(1). The spin-polarized ferromagnet (FM) and the fully layer
polarized (FLP) phases are symmetric, while the canted antiferromagnet
(CAF), the Kekule (KEK), and the spin-valley entangled (SVE) phases
break a single U(1) symmetry.

We explored three parameter sets of couplings characterized by
inequalities among them. For parameter set 1 (PS1), $g_z>g_0+\half
g_{nz}+|g_{xy}|$, and the BU(1)$^2$ phase invariably appears in the
$B_\perp-D$ phase diagram at small $B_\perp$ and small $D$ when the
$\bB$-field is not tilted. In this regime, transitions between the
CAF, BU(1)$^2$, SVE, and FLP phases are driven by increasing $D$ and
are all second-order. At large $B_\perp$ a partially orbitally
polarized (POP) phase, and the Kekul\'e (KEK) phase intervene between
the CAF and the FLP phases for intermediate values of $D$. Transitions
between the POP and other states are always first-order, while the
transition from the KEK state to the FLP state is second-order. As the
field is tilted and the Zeeman energy increased, the BU(1)$^2$ phase
shrinks and disappears from the $B_\perp-D$ phase diagram.

Parameter set 2 (PS2) satisfies the inequalities $g_0+\half
g_{nz}+|g_{xy}|>g_z>g_0+\half g_{nz}$. In this case the BU(1)$^2$
phase, if it appears at all, is confined to a small sliver of $D$ and
$B_\perp$ near the onset of the POP state when the $\bB$-field is
untilted. At small $B_\perp$ the CAF state transitions directly to the
SVE state via a first-order transition as $D$ is increased, which in
turn smoothly goes over into the FLP state via a second-order
transition at even higher $D$. As above, at larger $B_\perp$, the POP
state intervenes at intermediate $D$, and a KEK state may appear at
higher $D$ which ultimately gives way to the FLP state.

Parameter set 3 (PS3) satisfies $g_z <g_0+\half g_{nz}$, and
has the simplest phase diagram of all. The CAF/FM state at small $D$
undergoes a first-order transition to either the FLP or the POP
state, depending on the value of $B_\perp$. All transitions in PS3 are
first-order.

The BU(1)$^2$ phase, if it exists, always appears in a narrow window
of $D$. Since it undergoes second-order phase transitions to states
with a single broken U(1) at its $D$-boundaries, one (pseudo)spin-stiffness
must always vanish at each transition. In previous work we have shown
that in such cases the gap to edge transport vanishes at the
transition. Depending on the details of the stiffnesses, and the
temperature at which measurements are made, the BU(1)$^2$ phase may
appear to be metallic. An alternative possibility is that
quantum fluctuations disorder at least one of the broken U(1)'s to
form a symmetric phase with vanishing gap at either $D$-boundary.

Our results also raise a host of interesting questions. Foremost among
them is the issue of edge conduction in the various states. The BLG
edge is expected to break all lattice symmetries, but preserve
spin-rotation symmetry, because spin-orbit coupling is tiny. For the
CAF state in monolayer graphene the present authors showed that edge
conduction occurs via topological vortex excitations of the CAF order
parameter bound to an image antivortex near the edge
\cite{us_2014,us_2016}.  In a quantum Hall state such topological
objects carry charge due to the spin-charge relation\cite{QHFM}. In
BLG, the SVE and KEK states are valley analogues of the CAF, and it
remains to be seen whether this edge physics carries over to the two
latter phases. Perhaps the most interesting is the edge BU(1)$^2$
phase, because the bulk supports several flavors of topological
excitations (vortices can be formed from either of the two broken
U(1)'s). The effects of thermal and/or quantum disordering of the
BU(1)$^2$ state should also be explored.

Another set of interesting questions concerns fillings close to
$\nu=0$, particularly in the range $-4\le \nu\le 4$. All these
fillings nominally involve only the nearly degenerate set of Landau
levels around the Fermi energy for undoped BLG. Trigonal warping
likely impacts the phase diagram at such fillings, and a detailed
investigation could help identify the appropriate interaction regime
for BLG.  Lastly, on the theoretical side, a full
renormalization-group analysis for the short-range couplings in the
presence of $t_3$ and a quantizing magnetic field, while challenging,
would in principle indicate the scale of couplings that apply to
models such as we have analyzed, in which the degrees of freedom are
projected to a small number of Landau levels.

There are also intriguing connections between the phase transitions in
BLG at $\nu=0$ and recent ideas of critical deconfinement
\cite{senthil}, which is the phenomenon whereby the emergent degrees
of freedom at a phase transition are fractionalized in terms of the
order parameter fields on either side of the transition. The canonical
example of critical deconfinement is the Neel to Valence Bond Solid
transition in a class of two-dimensional quantum
antiferromagnets. Recall that in the absence of Zeeman coupling, the
CAF state would become an antiferromagnet (AF). Recently, it was
argued\cite{Lee-Sachdev_2015} that the transition between the AF and
the KEK phase would be critically deconfined. Adding the Zeeman
coupling will convert the deconfined transition into a region where
the two order parameters coexist\cite{senthil}. The BU(1)$^2$ phase
does have both CAF and KEK order parameters but, in our model, exists
even at zero Zeeman coupling. 

Last, but not least, it has been proposed\cite{xu-etal} that the fully
polarized FM state in BLG (achieved at large Zeeman coupling) could be
a realization of a bosonic symmetry-protected topological
insulator\cite{lu-vishwanath}. Precisely what set of interaction
parameters would realize such a state remains an open question.

\acknowledgements

 We are grateful to Jun Zhu, Jing Li, Andrea Young, Mike Zaletel, and
 Juan Ramon de Nova for illuminating conversations, and to the Aspen
 Center for Physics (NSF Grant 1066293), where this work was begun and
 completed. GM thanks the NSF (DMR-1306897) and the Gordon and Betty
 Moore Foundation for support.  HAF acknowledges the support of the
 NSF through grant Nos. DMR-1506263 and DMR-1506460. ES thanks support
 of the Israel Science Foundation (ISF) via grant no. 231/14, of the
 Simons Foundation, and thanks the hospitaliy of the Kavli Institute
 for Theoretical Physics (NSF PHY-11-25915). Finally we would like to
 acknowledge support for all the present authors by the US-Israel
 Binational Science Foundation (BSF-2012120).

\bigskip

\appendix

\section{Derivation of the coefficients $A_{nm}$}
\label{Anm}

In this Appendix we derive a power-series expansion in $\lambda$ for the states $|\psi_A\rangle$, $|\psi_B\rangle$ [Eq. (\ref{psi_AB_integral})], and consequently the expressions for the coefficients $A_{nm}$ in Eq. (\ref{psi_01_def}). We start by considering the integral
\beqr
\label{t_to_xi}
& &\int_0^\infty dt\,e^{\frac{it^3}{3\lambda}-ita^\dagger} = \\
& &\left(\frac{\lambda}{9}\right)^{1/3}e^{\frac{i\pi}{6}}\int_0^\infty d\xi\,\xi^{-2/3}e^{-\xi}\exp\left\{e^{-\frac{i\pi}{3}}(3\lambda\xi)^{1/3}a^\dagger\right\}\nonumber
\eeqr
where we have used the change of variables $t^3=i3\lambda\xi$. Implementing a power-series expansion of the last exponential factor in Eq. (\ref{t_to_xi}), and performing the integration over $\xi$, we obtain
\begin{widetext}
\beq
\label{int_to_series}
\int_0^\infty dt\,e^{\frac{it^3}{3\lambda}-ita^\dagger} =
\left(\frac{\lambda}{9}\right)^{1/3}e^{\frac{i\pi}{6}}\sum_{n=0}^\infty\frac{(3\lambda)^{n/3}\Gamma\left(\frac{n+1}{3}\right)}{n!}
e^{-\frac{in\pi}{3}}(a^\dagger)^n\; .
\eeq
Employing Eq. (\ref{psi_AB_integral}), we thus find
\beq
\label{psi_A_series}
|\psi_A\rangle=
\left(\frac{\lambda}{9}\right)^{1/3}\sum_{n=0}^\infty\frac{(3\lambda)^{n/3}\Gamma\left(\frac{n+1}{3}\right)}{n!}\cos\left\{\frac{\pi}{6}(2n-1)\right\}
(a^\dagger)^n|0\rangle\; .
\eeq
To get a similar expansion for $|\psi_B\rangle$, we repeat the same steps for the purely real integral
\beq
\label{realint_to_series}
\int_0^\infty dt\,e^{-\frac{t^3}{3\lambda}-ta^\dagger} =
\left(\frac{\lambda}{9}\right)^{1/3}\sum_{n=0}^\infty\frac{(3\lambda)^{n/3}\Gamma\left(\frac{n+1}{3}\right)}{n!}
(-1)^n(a^\dagger)^n\; ;
\eeq
substituting in Eq. (\ref{psi_AB_integral}), this yields
\beq
\label{psi_B_series}
|\psi_B\rangle=
\left(\frac{\lambda}{9}\right)^{1/3}\sum_{n=0}^\infty\frac{(3\lambda)^{n/3}\Gamma\left(\frac{n+1}{3}\right)}{n!}
\left[(-1)^n-\sin\left\{\frac{\pi}{6}(2n-1)\right\}\right]
(a^\dagger)^n|0\rangle\; .
\eeq

We next examine the oscillating factors in Eqs. (\ref{psi_A_series}) and (\ref{psi_B_series}), which exhibit a 3-fold periodicity in $n$: for any integer $m$,
\beqr
n &=& 3m-1\quad\Rightarrow\quad \cos\left\{\frac{\pi}{6}(2n-1)\right\}=(-1)^n-\sin\left\{\frac{\pi}{6}(2n-1)\right\}=0 \nonumber\\
n &=& 3m\quad\Rightarrow\quad \cos\left\{\frac{\pi}{6}(2n-1)\right\}=(-1)^m\frac{\sqrt{3}}{2}\; ,\quad(-1)^n-\sin\left\{\frac{\pi}{6}(2n-1)\right\}=(-1)^m\frac{3}{2} \nonumber\\
n &=& 3m+1\quad\Rightarrow\quad \cos\left\{\frac{\pi}{6}(2n-1)\right\}=(-1)^m\frac{\sqrt{3}}{2}\; ,\quad (-1)^n-\sin\left\{\frac{\pi}{6}(2n-1)\right\}=(-1)^{m+1}\frac{3}{2}\; .
\label{osc_factors}
\eeqr
Inserting Eq. (\ref{osc_factors}) in (\ref{psi_A_series}), (\ref{psi_B_series}) and using $|N\rangle=\frac{1}{\sqrt{N!}}(a^\dagger)^N|0\rangle$, we obtain
\beqr
|\psi_A\rangle &=& \frac{1}{\sqrt{3}}\left(|\tilde{\psi}_0\rangle+|\tilde{\psi}_1\rangle\right) \nonumber \\
|\psi_B\rangle &=& |\tilde{\psi}_0\rangle-|\tilde{\psi}_1\rangle
\eeqr
where
\beqr
|\tilde{\psi}_0\rangle &=& \frac{(3\lambda)^{1/3}}{2}\sum_{m=0}^\infty (-1)^m\frac{(3\lambda)^m}{\sqrt{(3m)!}}\Gamma\left(m+\frac{1}{3}\right) |3m\rangle  \nonumber \\
|\tilde{\psi}_1\rangle &=& \frac{(3\lambda)^{1/3}}{2}\sum_{m=0}^\infty (-1)^m\frac{(3\lambda)^m}{\sqrt{(3m+1)!}}\Gamma\left(m+\frac{2}{3}\right) |3m+1\rangle\; .
\label{psi_01_unnorm}
\eeqr
\end{widetext}
By definition, $|\tilde{\psi}_n\rangle$ are orthogonal ($\langle\tilde{\psi}_0 |\tilde{\psi}_1\rangle$=0) for arbitrary prefactors of each. Hence, introducing the normalization factors $C_0$, $C_1$, we arrive at the orthonormal basis states Eq. (\ref{psi_01_def}). Once this form has been obtained, it is straightforward to verify that these states satisfy $(a^2+\lambda a^{\dagger})|\tilde{\psi}_0\rangle=0$.

\section{Form Factors}
\label{app:FF}
In this Appendix we discuss some details relevant to the calculation of the
density matrix elements, Eq. (\ref{eq:trho}), and in particular how their form
leads to Eq. (\ref{eq:trho_integral}).  We begin with the basis states $|n,\alpha,k\rangle$
in Eq. (\ref{nalpha_def}),
\begin{widetext}
\beq
|n,\alpha,k\rangle\equiv\sum_{m=0}^\infty (-1)^{m\alpha}A_{nm}|3m+n,k\rangle\; ,
\eeq
for which the coefficients $A_{nm}$ are defined in Eq. (\ref{psi_01_def}).
Direct substitution yields the explicit form
\begin{equation}
\tilde\rho^{\alpha\beta}_{n_1n_2}({\bf q})=
\sum_{k_1=0}^{\infty}\sum_{k_2=0}^{\infty}(-1)^{k_1\alpha+k_2\beta}
A_{n_1k_1}A_{n_2k_2}\rho_{3k_1+n_1,3k_2+n_2}(\bf{q})
\end{equation}
where the usual Landau level matrix elements are defined as 
\begin{equation}
\rho_{n_1n_2}({\bf q})=(-1)^{n_<+n_2}e^{-q^2\ell^2/4} \sqrt{\frac{n_{<}!}{n_{>}!}}
\,e^{i(n_1-n_2)(\theta_q-\pi/2)} \left(\frac{q\ell}{\sqrt{2}}\right)^{n_{>}-n_{<}}
L_{n_<}^{|n_{1}-n_{2}|}\left(\frac{q^2\ell^2}{2}\right).
\end{equation}
In this equation, $n_{<}$ ($n_{>}$) is the smaller (larger) of $n_1$ and $n_2$, $L_m^n$ is an
associated Laguerre polynomial, and $\theta_q$ is the angle formed by ${\bf q}$ with
the $\hat{x}$-axis.  Now consider the exchange integral 
\begin{eqnarray}
&\int\frac{d^2q}{(2\pi)^2}&\!\!\!\!\!\!\!\!\!\!
v({\bf q})
\tilde\rho_{n_1n_2}^{\alpha\beta}({\bf q})\tilde\rho_{m_1m_2}^{\gamma\delta}(-{\bf q}) \nonumber \\
&=\sum_{k_1k_2k_3k_4}&(-1)^{k_1\alpha+k_2\beta+k_3\gamma+k_4\delta}
A_{n_1k_1}A_{n_2k_2}A_{m_1k_3}A_{m_2k_4}
\int \frac{d^2q}{(2\pi)^2} v({\bf q})
\rho_{3k_1+n_1,3k_2+n_2}({\bf q})\rho_{3k_3+m_1,3k_4+m_2}(-{\bf q}).\nonumber\\
\label{rhorhoint}
\end{eqnarray}
Writing $N_1\equiv 3k_1+n_1$, $N_2\equiv 3k_2+n_2$, $M_1\equiv 3k_3+m_1,$ and
$M_2\equiv 3k_4+m_2,$ Eq. (\ref{rhorhoint}) can be reexpressed as
\begin{eqnarray}
\int\frac{d^2q}{(2\pi)^2}v({\bf q})
\tilde\rho_{n_1n_2}^{\alpha\beta}({\bf q})\tilde\rho_{m_1m_2}^{\gamma\delta}(-{\bf q})
=\sum_{k_1k_2k_3k_4}(-1)^{k_1\alpha+k_2\beta+k_3\gamma+k_4\delta}
A_{n_1k_1}A_{n_2k_2}A_{m_1k_3}A_{m_2k_4} \nonumber\\
\times \, \int \frac{d^2q}{(2\pi)^2} v({\bf q}) e^{-q^2\ell^2/2}
(-1)^{N_{<}+N_2+M_<+M_2+M_1+M_2}\sqrt{\frac{N_<!M_<!}{N_>!M_>!}}
\,e^{i(\theta_1-\frac{\pi}{2})(N_1-N_2+M_1-M_2)} \nonumber\\
\times\,\left(\frac{q\ell}{\sqrt{2}}\right)^{|N_1-N_2|+|M_1-M_2|}
L_{N_<}^{|N_1-N_2|}\left(\frac{q^2\ell^2}{2}\right)
L_{M_<}^{|M_1-M_2|}\left(\frac{q^2\ell^2}{2}\right),
\label{B5}
\end{eqnarray}
\end{widetext}
where we used the property $$\rho_{n_1n_2}(-{\bf q})=(-1)^{n_1+n_2}\rho_{n_1n_2}({\bf q}).$$
The integration over $\theta_q$ forces the integral to vanish unless
$N_1+M_1=N_2+M_2$.  Moreover, specializing to the case where $v({\bf q})$
has no ${\bf q}$ dependence, the orthogonality relation
\begin{equation}
\int_0^\infty dx \, e^{-x}x^{\alpha} L_m^{\alpha}(x) L_n^{\alpha}(x) = \frac{\Gamma(n+\alpha+1)}{n!}\delta_{mn}
\label{ortho_relation}
\end{equation}
guarantees that the integral in Eq. (\ref{B5}) vanishes unless $N_<=M_<$.
Writing $v({\bf q}) \rightarrow \tilde v$, we arrive at the relation
\begin{widetext}
\begin{equation}
\int\frac{d^2q}{(2\pi)^2}v({\bf q})
\tilde\rho_{n_1n_2}^{\alpha\beta}({\bf q})\tilde\rho_{m_1m_2}^{\gamma\delta}(-{\bf q})
= \frac{\tilde v}{2\pi \ell^2} \delta_{n_1m_2} \delta_{m_1n_2}
r_{\alpha\delta}^{(n_1)} r^{(n_2)}_{\beta\gamma}
\end{equation}
where
\begin{equation}
r_{\alpha\beta}^{(n)} = \sum_{k=0}^{\infty} (-1)^{k(\alpha+\beta)} A_{nk}^2
\equiv\delta_{\alpha\beta}+r(1-\delta_{\alpha\beta}).
\label{rabn}
\end{equation}
\end{widetext}
$r_{\alpha\beta}^{(n)}$ turns out to be unity if $\alpha = \beta$ because of the
normalization condition that the wavefunctions coefficients $A_{nk}$ must obey.
For $\alpha \ne \beta$, the sum is non-trivial, but we have found by direct summation that its value
is the {\it same} for both values of $n$ to within any numerical accuracy we can attain.
For this reason the quantity
$$
r = \sum_{k=0}^{\infty} (-1)^{k} A_{nk}^2
$$
is for all intents and purposes independent of $n$.  Eq. (\ref{rabn}) yields the result
used in Eq. (\ref{eq:trho_integral}).

\end{document}